\documentclass[manuscript,doublespace]{aastex61}
\hypersetup{linkcolor=red,citecolor=blue,filecolor=cyan,urlcolor=magenta}

\newcommand{\Alfven}{Alfv\'en}

\submitjournal{ApJ}
\shorttitle{Stratified collisionless disks simulations}
\shortauthors{Hirabayashi et al.}

\begin{document}

\title{Stratified Simulations of Collisionless Accretion Disks}

\correspondingauthor{Kota Hirabayashi}
\email{hirabayashi-k@eps.s.u-tokyo.ac.jp}

\author{Kota Hirabayashi}
\affiliation{Department of Earth and Planetary Science,
The University of Tokyo, Tokyo, 113-0033, Japan}

\author{Masahiro Hoshino}
\affiliation{Department of Earth and Planetary Science,
The University of Tokyo, Tokyo, 113-0033, Japan}

% \author{Takanobu Amano}
% \affiliation{Department of Earth and Planetary Science,
% The University of Tokyo, Tokyo, 113-0033, Japan}

\begin{abstract} % single paragraph, < 251 words
 This paper presents a series of stratified shearing-box simulations of
 collisionless accretion disks in the recently developed framework of
 kinetic magnetohydrodynamics (MHD), which can handle finite
 non-gyrotropy of a pressure tensor.
 Although a fully kinetic simulation predicted a more efficient
 angular-momentum transport in collisionless disks than in the standard
 MHD regime, the enhanced transport has not been observed in past
 kinetic MHD approaches to gyrotropic pressure anisotropy.
 For the purpose of investigating this missing link between the fully
 kinetic and MHD treatments, this paper pays attention to the role of
 non-gyrotropic pressure, and makes a first attempt to incorporate
 certain collisionless effects into disk-scale, stratified disk
 simulations.
 When the timescale of gyrotropization was longer than, or comparable
 to, the disk rotation frequency of the orbit, we found that the finite
 non-gyrotropy selectively remaining in the vicinity of current sheets
 contributes to suppressing magnetic reconnection in the shearing-box
 system.
 This leads to increases both in the saturated amplitude of the MHD
 turbulence driven by magnetorotational instabilities and in the
 resultant efficiency of angular-momentum transport.
 Our results seem favorable for fast advection of magnetic fields toward
 the rotation axis of a central object, which is required to launch an
 ultra-relativistic jet from a black-hole accretion system in, for
 example, a magnetically arrested disk state.
 % (222 words)
\end{abstract}

\keywords{accretion, accretion disks --- plasmas ---
magnetohydrodynamics (MHD) --- methods: numerical}

 \section{Introduction}
 \label{sec:introduction}
 Explaining an efficient angular-momentum transport required for mass
 accretion in accretion disks is a fundamental issue in astrophysics.
 Since the astrophysical importance of magnetorotational instability
 (MRI) was pointed out \citep{1991ApJ...376..214B}, MRI has been
 investigated elaborately as a driver of strong magnetohydrodynamic
 (MHD) turbulence to provide a substantial turbulent transport of
 angular momentum.
 Although a number of numerical studies of MRIs developed under the MHD
 framework, in which a collisional state is assumed, have achieved
 success,
 % add refefrence here
 % \citep[e.g.,][]{},
 it is also of great importance to understand the dynamics in the
 collisionless regime.
 For instance, Sagittarius A* (Sgr A*), which is a compact radio source
 at the center of our galaxy, is believed to be a combination of a
 collisionless accretion disk and a relativistic jet possessing a
 supermassive black hole at its center
 \citep[e.g.,][]{1995Natur.374..623N,2000A&A...362..113F,2008Natur.455...78D,2012ApJ...748...34K}.

 Given this fact, \cite{2015PhRvL.114f1101H} conducted a
 three-dimensional, local-shearing-box simulation in the collisionless
 regime using a particle-in-cell (PIC) technique.
 He pointed out that the angular-momentum transport carried by
 anisotropy in the velocity-distribution function, which is interpreted as
 an anisotropic pressure tensor in fluid-based models, reached a
 value comparable to that carried by the Maxwell stress.
 It was also argued that the total transport efficiency measured by
 instantaneous thermal pressure was enhanced by an order of magnitude
 compared with standard MHD results.
 In terms of the $\alpha$-viscosity proposed in
 \cite{1973A&A....24..337S}, which is defined as the $xy$-component of
 the stress tensor normalized by the thermal pressure,
 $\alpha \sim \mathcal{O}\left(0.1\right)$ was achieved.

 \cite{2006ApJ...637..952S}, on the other hand, investigated the effect
 of the anisotropic stress under the fluid framework based on a
 combination of the double-adiabatic approximation, or
 Chew-Goldberger-Low (CGL) model \citep{1956RSPSA.236..112C}, and the
 Landau fluid model \citep{1990PhRvL..64.3019H}, where a gyrotropic
 pressure is assumed.
 Gyrotropy means that a pressure tensor can be described by only two
 independent components parallel and perpendicular to a local magnetic
 field, rather than one scalar value.
 These two variables evolve with time in association with MHD-type
 equations, and the other kinetic effects are neglected. 
 The CGL results agree with the more lately published PIC calculation
 in the sense that the anisotropic and Maxwell stresses had the same
 contribution to angular-momentum transport. 
 A puzzling difference between these two approaches is, however, the
 total efficiency of the transport.
 Although the PIC simulation predicts a quite large value of $\alpha$,
 the local simulations using the CGL model lead to values similar to
 those obtained in standard MHD calculations.

 One possible key to solving this discrepancy may be the treatment of
 magnetic reconnection under anisotropic pressure.
 In both fully kinetic and fluid-based approaches, the perpendicular
 pressure, $p_{\perp}$, tends to dominate the parallel pressure,
 $p_{||}$, in shearing-box simulations.
 This is a qualitative result of the fact that MRI is a process for
 enhancing the magnetic field, which naturally results in anisotropy
 with $p_{\perp}>p_{||}$ from conservation of the first and second
 adiabatic invariants.
 In contrast, magnetic reconnection dissipates the magnetic energy and
 causes opposite anisotropy in neutral sheets where reconnection takes
 place.
 According to \cite{2015PhRvL.114f1101H}, it is this parallel-pressure
 enhancement via reconnection that suppresses successive reconnection
 itself \citep{1984PhFl...27.1198C}.
 As a result, a larger magnetic energy, and hence larger Maxwell stress,
 is likely to be maintained in the system.
 It is, however, still ambiguous whether and to what extent the effect
 of suppression of reconnection by the pressure anisotropy is retained
 in the CGL framework
 \citep[e.g.,][]{2001JGR...106.3737B,2016JGRA..121.6245H}.
 A part of this ambiguity arises from the fact that CGL-based models
 cannot deal with pressure anisotropy in magnetically neutral regions in
 its own right owing to the presence of the singularity near $B=0$.
 To take a step forward beyond previous fluid models with anisotropy, in
 this paper, we adopt a model recently developed by
 \cite{2016JCoPh.327..851H} (hereafter HHA16).
 This model enables us to define and resolve the anisotropic pressure
 tensor both in magnetized and unmagnetized regions seamlessly without
 any peculiar treatment by introducing a timescale of gyrotropization.
 Thus, we can investigate the role of the pressure tensor at the deep
 inside the neutral sheets.

 In spite of the limited kinetic effects included in the system, it is
 necessary to employ a fluid-based model possessing the scale-free
 property, when large-scale dynamics are considered.
 Since the fully kinetic approach must resolve particle scales such as
 Debye lengths and gyro radii, it is quite unrealistic to analyze
 disk-scale behavior, which, in general, occurs on a scale much larger
 than the kinetic one.
 In this sense, this work can be considered as one of the attempts to
 bridge the gap between the small-scale kinetic approach and the large-scale
 fluid approach.
 In particular, we tackle a series of {\it stratified}-shearing-box
 simulations by retaining the vertical gravity of a central object.
 The stratification introduces the concept of the disk's scale height or
 the disk thickness into the system.
   Due to the limitation of present computational power, this scale
   apparently cannot be reproduced in PIC simulations where all kinetic
   scales must be resolved simultaneously.

 The differences between stratified- and unstratified-shearing-box
 simulations have been discussed by a number of authors, based on the
 standard MHD framework in the collisional regime
 \citep[e.g.,][]{1995ApJ...446..741B,1996ApJ...463..656S,2000ApJ...534..398M,2010ApJ...713...52D}.
 One major change is generation of buoyantly rising patches of strong
 toroidal magnetic fields, even when the simulation domain does not
 contain any external magnetic flux.
 Previous studies have suggested that these magnetic patches work
 just like a global or an external field for local MRIs, which can
 sustain strong MRI-driven turbulence and enhance the angular-momentum
 transport by roughly one order of magnitude in terms of $\alpha$
 compared with an unstratified simulation starting from the same initial
 condition.
 It is still unclear what makes these patches.
 Nevertheless, investigating this effect in the collisionless regime as
 well would be of importance to understanding the global behavior of
 collisionless accretion disks.
 In the present paper, we revisit this point for the first time in the
 collisionless framework.
 
 This paper is organized as follows.
 In Section \ref{sec:setups}, we describe the numerical settings.
 In particular, procedures specific to the present problem are explained
 in detail.
 Section \ref{sec:results} discusses our simulation results in
 comparison to cases under isotropic pressure calculated using the same
 code.
 Finally, Section \ref{sec:summary} is devoted to summary and
 concluding remarks. 
 
 \section{Simulation setup}
 \label{sec:setups}

  \subsection{Basic equations}
  \label{sec:equations}
  In all calculations presented in this paper, we employ the
  local-shearing-box approximation \citep{1995ApJ...440..742H}
  incorporated with the vertical component of gravity.
  The pressure anisotropy is handled with the model described in HHA16,
  and the shearing source terms and Coriolis force are additionally
  included.
  Then, the basic equations are as follows:
  \begin{eqnarray}
   \frac{\partial\rho}{\partial t}
   + \nabla \cdot \left(\rho\mathbf{v}\right) = 0,
   \label{eq:eq_mass} \\
   \frac{\partial\rho\mathbf{v}}{\partial t}
    + \nabla \cdot
    \left( \rho\mathbf{vv} + \mathbf{p}
     + \frac{B^2}{2} \mathbf{I} - \mathbf{BB} \right) =
    - 2 \rho \mathbf{\Omega} \times \mathbf{v}
    - \rho \nabla \Phi,
    \label{eq:eq_mom} \\
   \frac{\partial\mathbf{B}}{\partial t}
    - \nabla \times \left(\mathbf{v}\times\mathbf{B}\right) = 0,
    \label{eq:eq_induction} \\
   \partial_t
    \left(\rho v_i v_j + p_{ij} + B_i B_j\right)
    + \partial_k
    \left(\rho v_i v_j v_k + p_{ij} v_k + p_{ik} v_j + p_{jk} v_i
     + \mathcal{S}_{kij} + \mathcal{S}_{kji}\right) \nonumber \\
   =
    B_i v_k \partial_k B_j + B_j v_k \partial_k B_i
    - B_k v_i \partial_j B_k - B_k v_j \partial_i B_k
   \nonumber \\
   - 2 \rho v_i \varepsilon_{jkl} \Omega_k v_l
   - 2 \rho v_j \varepsilon_{ikl} \Omega_k v_l
   - \rho v_i \partial_j \Phi - \rho v_j \partial_i \Phi
   - \nu_g\left(p_{ij} - p_{g,ij}\right),
   \label{eq:eq_energy}
  \end{eqnarray}
  where $\mathbf{\Omega}=\Omega\hat{\mathbf{e}}_z$ is an angular
  velocity vector,
  $\mathcal{S}_{kij} = c\varepsilon_{kli}E_l B_j$
  is a generalized Poynting flux tensor, $\varepsilon_{ijk}$ is the
  Levi-Civita symbol, $\Phi$ is a gravitational potential, and the other
  notations are standard.
  Note that the factor $1/\sqrt{4\pi}$ is absorbed into the definition
  of the magnetic field.
  Throughout this paper, we assume the ideal Ohm's law,
  $\mathbf{E}+\left(\mathbf{v}/c\right)\times\mathbf{B}=0$.
  The gravitational potential $\Phi$ in the shearing box with vertical
  gravity is written as
  \begin{eqnarray}
   \Phi = - q \Omega^2 x^2 + \frac{1}{2} \Omega^2 z^2,
  \end{eqnarray}
  where $q=-d\ln\Omega/d\ln R$ is a shear parameter and we assume
  $q=3/2$, which corresponds to Keplerian rotation.

  To mimic pitch angle scattering due to micro-instabilities driven by
  the pressure anisotropy, we adopt the so-called hard-wall limit
  introduced in \cite{2006ApJ...637..952S}.
  This model sets a maximum extent of the anisotropy beyond which the
  pressure tensor is isotropized immediately to the marginal value.
  The detailed implementation of isotropization via the scattering model
  in our code is performed following the technique explained in HHA16.

  In addition to gyrotropization and isotropization, we implement the
  effect of cooling, or {\it isothermalization}, in a rather artificial
  manner.
  This step is required to maintain the vertical structure of the disk;
  without any cooling, the gas in a simulation box is heated
  continuously, making the disk thicker and thicker.
  For the purpose of investigating long-term evolution under a
  statistically constant disk structure, therefore, the dissipated
  thermal energy must be removed from the computational domain.
  To achieve this, we employ a technique similar to the
  gyrotropization model.
  Once $\rho$ and $p_{ij}$ are found from Equations (\ref{eq:eq_mass})
  and (\ref{eq:eq_energy}), respectively, we can define the
  {\it isothermal} pressure tensor,
  \begin{eqnarray}
   \mathbf{p}_{cool} = \rho c_s^2 \
    \frac{3\mathbf{p}}{p_{xx}+p_{yy}+p_{zz}},
    \label{eq:cooling}
  \end{eqnarray}
  where $c_s$ is the speed of sound, which is assumed to be uniform and
  constant.
  Equation (\ref{eq:cooling}) renormalizes the pressure tensor such that
  the diagonal average of $\mathbf{p}$, which is directly proportional
  to the total thermal energy, becomes identical to the isothermal
  value $\rho c_s^2$, while the ratios between every two components of
  the pressure tensor remain unchanged.
  Then, the pressure is made to approach $\mathbf{p}_{cool}$ nearly
  instantaneously as
  \begin{eqnarray}
   \left.
    \frac{\partial\mathbf{p}}{\partial t}
   \right|_{cool} =
    - \nu_{cool}
    \left(\mathbf{p}-\mathbf{p}_{cool}\right),
  \end{eqnarray}
  with $\nu_{cool}$ assumed to be a large number compared to, for
  example, the disk-rotation frequency $\Omega$.

    Note that, in order to avoid explicit integration of a stiff equation,
    various relaxation terms with quite large frequencies are treated
    separately in an operator splitting manner with the help of an analytic
    solution of each effect, as in HHA16.

  \subsection{Initial and boundary conditions}
  \label{sec:conditions}
  Equations (\ref{eq:eq_mass})--(\ref{eq:eq_energy}), in combination
  with the isotropization and cooling models, are solved as an initial-
  and boundary-value problem.
  Initially, we assume a purely azimuthal differential rotation expanded
  linearly in the local shearing box,
  \begin{eqnarray}
   \mathbf{v}_K = - q \Omega x \hat{\mathbf{e}}_y.
  \end{eqnarray}
  Suppose that the gas pressure is isotropic at $t=0$.
  The stratified disk is then given by a vertical hydrostatic
  balance.
  Using the isothermal relation, we obtain
  \begin{eqnarray}
   \rho =
    \rho_0 \exp \left(-z^2/H^2\right),  \\
   \mathbf{p} =
    p_0 \mathbf{I} \exp \left(-z^2/H^2\right),
  \end{eqnarray}
  where $\rho_0$ is the mid-plane density, $p_0=\rho_0 c_s^2$ is the
  mid-plane pressure, and $H=\sqrt{2}c_s/\Omega$ is the disk thickness.
  A magnetic field is imposed upon this disk structure.
  In this paper, the sinusoidally changing vertical field is
  considered as
  \begin{eqnarray}
   \mathbf{B} =
    B_0 \sin \left(\frac{2\pi x}{L_x}\right) \hat{\mathbf{e}}_z,
  \end{eqnarray}
  where $B_0$ is the maximum field strength and $L_x$ is the radial
  dimension of the simulation domain.
  It is known that the intensity of MRI-driven turbulence is highly
  sensitive to the net vertical magnetic flux, which vanishes in the
  present configuration.
  This zero-net-flux model is motivated by the fact that there is no
  obvious observational evidence that black-hole-accretion flows are
  threaded by large-scale external magnetic flux.

  The above initial configuration is in magnetohydrodynamical
  equilibrium.
  To seed the growth of MRIs, we superpose a random, isentropic
  perturbation on the density and gas pressure, the amplitude of
  which is 0.1\% of the local background value.
  For consistency with \cite{1995ApJ...440..742H}, a velocity
  perturbation is also added with an amplitude of 0.02\% of $c_s$.

  We adopt a shearing periodic boundary condition in the $x$-direction
  while the effect of the background shear is taken into account
  \citep{1995ApJ...440..742H,2010ApJS..189..142S}.
  In practice, this is achieved by a data shift along the azimuthal flow
  after the periodic boundary condition is applied.
  In the $y$-direction, a purely periodic boundary is employed.

  A particular issue with stratified-disk simulations comes into the
  discussion when the vertical boundary is considered.
  We assume a periodic boundary in the $z$-direction, although this
  appears to be unrealistic in the present stratified box.
  The extent to which the employment of the periodic vertical boundary
  affects the disk dynamics has been discussed by several authors.
  \cite{1996ApJ...463..656S} compares two runs: one adopting a
  periodic boundary condition and the other an outflow boundary
  condition with an additional vertical domain including strong
  viscosity and resistivity.
  This extended region works as a dumping layer, which is necessary to
  avoid a spurious Lorentz force arising from an artificially snipped
  magnetic field line at the boundary.
  They concluded that neither vertical structure nor volume-averaged
  quantities show any significant difference between the two runs.
  \cite{2010ApJ...713...52D}, on the other hand, performed simulations
  with vertical dimensions of $4H$ and $6H$ using a periodic boundary.
  They also concluded that the nature of MRI turbulece and the
  angular-momentum transport in both cases looks quite similar, and
  hence, the four-scale-height run associated with the periodic boundary
  yields robust estimates of the properties of disk turbulence.
  In our implementation, the gravitational potential is modified to
  connect the values at the top and bottom boundaries smoothly, as
  described in \cite{2010ApJ...713...52D}, i.e.,
  \begin{eqnarray}
   \frac{1}{2} \Omega^2 z^2 \to
    \frac{1}{2} \Omega^2
    \left(
     \sqrt{\left(z_0-\left|z\right|\right)^2 + \lambda^2} - z_0
    \right)^2,
  \end{eqnarray}
  where $z_0$ indicates the position of the top or bottom boundary, and
  $\lambda$ is the thickness of the smoothing region.
  We set $\lambda=0.1H$.
  
  \subsection{Code}
  \label{sec:code}
  The code that we use is a higher-order version of that described in
  HHA16, namely, fifth-order accuracy in space and third-order accuracy
  in time.
  The gyrotropization, isotropization, and cooling procedures are
  combined in an operator-splitting manner.

  For the purpose of reducing and homogenizing errors that arise from
  the background shearing velocity, we employ the orbital-advection
  technique introduced in \cite{2010ApJS..189..142S}.
  This scheme decomposes Equations
  (\ref{eq:eq_mass})--(\ref{eq:eq_energy}) into two systems by making
  use of the fact that the background shear-flow velocity $\mathbf{v}_K$
  is constant in time.
  One is the usual MHD system where
  $\mathbf{v}^{\prime}=\mathbf{v}-\mathbf{v}_K$
  is evolved rather than $\mathbf{v}$ itself, with a slight modification
  of the shearing source terms.
  The other system describes the linear advection due to the
  background shear flow, which can be solved analytically.
  See \cite{2010ApJS..189..142S} for the technical details developed in
  standard MHD.
  For completeness sake, implementation in the present model with an
  anisotropic pressure is summarized in Appendix~\ref{app:advection}.

 \section{Results}
 \label{sec:results}
 Important parameters and spatially and temporally averaged stresses in
 our simulations are summarized in Table~\ref{tab:runs}.
 The three-dimensional simulation domain of the radial, azimuthal, and
 vertical coordinates is
 $\left(x,y,z\right) \in [-H/2, H/2] \times [0, 4H] \times [-2H, 2H]$
 in all cases.
 Each run is labeled as A (with anisotropic pressure) or I (with
 isotropic pressure), and 32 and 64 indicate the numbers of grid points
 per scale height, as shown in the second column.
 Runs A32 are further distinguished by lowercase alphabets.
 The third column is the simulation end time.
 The plasma beta and the gyrotropization frequency employed in each run
 are shown in the fourth and fifth columns, respectively.
 Throughout this paper, double brackets $\left<\left<f\right>\right>$
 represent spatial averages over the whole simulation domain and
 temporal averages after 50 orbits unless otherwise specified,
 whereas single brackets $\left<f\right>$ indicate only spatial averages.
 The averaged Reynolds, Maxwell, and anisotropic stresses normalized by
 the mid-plane pressure are recorded in the remaining columns.
 
 \begin{deluxetable}{cccccccc}
  \tablecaption{Simulation summary\label{tab:runs}}
  \tablehead{
  \colhead{Run} &
  \colhead{Resolution} &
  \colhead{Orbits} &
  \colhead{$\beta$} &
  \colhead{$\nu_{g0}/\Omega$} &
  \colhead{$\left<\left<\rho v_x v_y\right>\right>/p_0$} &
  \colhead{$\left<\left<-B_x B_y\right>\right>/p_0$} &
  \colhead{$\left<\left<p_{xy}\right>\right>/p_0$}
  }
  \colnumbers
  \startdata
  A32a & $32/H$ & 300 & $10^2$ & $10^{13}$ & 0.0014 & 0.0044 & 0.0020 \\
  A32b & $32/H$ & 100 & $10^3$ & $10^{13}$ & 0.0013 & 0.0040 & 0.0020 \\
  A32c & $32/H$ & 100 & $10^4$ & $10^{13}$ & 0.0015 & 0.0050 & 0.0019 \\
  A32d & $32/H$ & 300 & $10^2$ & $10^{8}$  & 0.0016 & 0.0049 & 0.0019 \\
  A32e & $32/H$ & 300 & $10^2$ & $10^{3}$  & 0.0016 & 0.0065 & 0.0020 \\
  A64  & $64/H$ &  20 & $10^2$ & $10^{13}$ & 0.0034 & 0.0124 & 0.0040 \\
  I32  & $32/H$ & 300 & $10^2$ & ---       & 0.0017 & 0.0062 & ---    \\
  I64  & $64/H$ &  20 & $10^2$ & ---       & 0.0018 & 0.0082 & ---    \\
  \enddata
  \tablecomments{Double brackets denote spatial averages over the whole
  simulation domain and temporal averages after 50 orbits, except for
  A64 and I64, where averages are taken after 10 orbits.}
 \end{deluxetable}

 Comparison among Runs A32a--A32c shows that the magnitude of the
 initial magnetic field does not affect the statistics in the later
 stage in either a qualitative or quantitative manner.
 This agreement essentially arises from the fact that, in zero-net-flux
 simulations, the system has no typical scale of magnetic flux.
 Once the initial magnetic structure is stirred by a non-linear growth
 of MRIs, therefore, information on the initial field is forgotten
 completely, and the systems tend to reach the same statistical steady
 state \citep{2006MNRAS.372..183P}.
 Then, we fix the value of the initial $\beta$ at 100 in other runs.

  \subsection{Fiducial run}
  \label{subsec:fiducial}
\begin{figure}[!ht]
   \centering
   \includegraphics[width=0.98\textwidth]{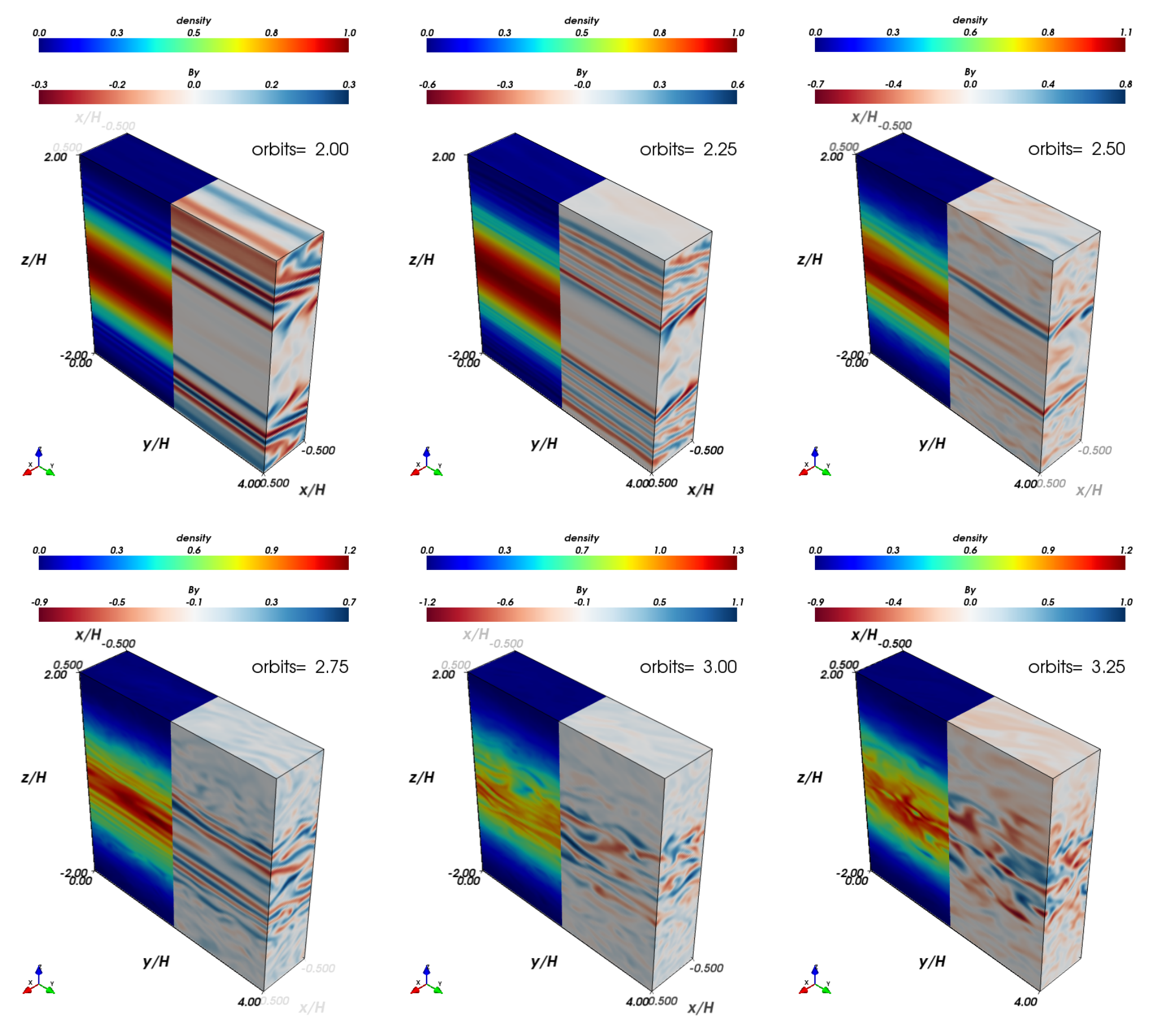}
   \caption{Snapshots during the initial phase of the MRI from Run A32a.
   The left and right halves in each panel show contours of the mass
   density and the azimuthal magnetic field, as normalized by the
   mid-plane density and the maximum field strength at the initial
   state, respectively.}
   \label{fig:mri_initial}
  \end{figure}

  We regard Run A32a as a fiducial case and review the result in this
  section.
  In Figure~\ref{fig:mri_initial}, color contours of the density (left
  half) and the azimuthal magnetic field (right half) normalized by
  $\rho_0$ and $B_0$, respectively, during the initial stage from 2 to
  3.25 orbits are shown with the time interval of 0.25 orbits.
  At first, the growth of the MRI becomes prominent in the outer regions
  with roughly $\left|z\right|>H$, although the linear theory in a
  uniform medium predicts that the maximum growth rate is $0.75\Omega$
  ubiquitously.
  As time goes on, MRI-driven channel sheets appear inside the disk as
  well.
  After 3 orbits, they finally break down into turbulence
  through magnetic reconnection, as observed in a number of previous
  studies using the standard MHD.
  \citep[e.g.,][]{1996ApJ...463..656S,2009ApJ...697.1269J,2009ApJ...691L..49S}.
  This compressible turbulence disturbs the disk structure, as seen in
  the density contours at 3 and 3.25 orbits, but the stratification is
  maintained statistically throughout the simulation, even in the
  presence of pressure anisotropy, which is discussed in detail later.
  
   \subsubsection{Statistics}
   After 3--4 orbits pass, the simulation box is filled with chaotic
   turbulent motion.
   Statistics of various quantities are summarized in
   Table~\ref{tab:statistics}.
   For comparative purpose, we also show the statistics for Run I32,
   where all conditions are the same as in Run A32a, except for the use
   of isotropic pressure.
   The second and fourth columns from the left represent averages both
   in space and in time, whereas the third and fifth columns show
   standard deviations of volume-averaged values, which indicate the
   magnitude of temporal fluctuation.
   
   From the energy distribution in Table~\ref{tab:statistics}, we can
   see that the properties of turbulent fluctuation in the two runs are
   quite similar.
   Magnetic energy, for example, is distributed into each directional
   component related to $B_x$, $B_y$, and $B_z$, respectively,
   roughly in the ratio of 0.1:0.85:0.05, and kinetic energy is
   $\sim 40\%$ of the total magnetic energy.
   The magnitudes of both the kinetic energy and magnetic energy,
   however, decrease by $\sim 20\%$ in A32a compared with the case in
   I32.
   A qualitatively similar relation also holds for stress;
   both Reynolds and Maxwell stresses are reduced by $\sim 20\%$ with
   a nearly constant ratio between them.
   In Run A32a, we also have additional angular-momentum transport by
   anisotropic stress, $p_{xy}$.
   The value of this anisotropic stress seems merely comparable to the
   contribution from the Reynolds stress.
   However, the 20\% reduction of other stress is compensated by
   $p_{xy}$, and the total transport efficiency does not change
   significantly.
   The bottommost two rows in Table~\ref{tab:statistics} compare the
   parallel and perpendicular pressures averaged over the simulation.
   As qualitatively predicted from the double-adiabatic approximation,
   in an average sense, $p_{\perp}$ dominates $p_{||}$, which causes
   positive $p_{xy}$ by combination with the positive Maxwell stress,
   because the anisotropic stress can be written as
   $p_{xy} \simeq \left(-B_x B_y/B^2\right) \left(p_{\perp} - p_{||}\right)$
   when the gyrotropic assumption holds.
   %\footnotemark[1]
   %\footnotetext[1]{Taking the $xy$-component of a gyrotropic pressure
   %tensor $\mathbf{p} = p_{\perp}\mathbf{I}+\left(p_{\parallel}-p_{\perp}\right)\mathbf{BB}/B^2$.}
   
   \begin{deluxetable}{lcccc}
    \tablecaption{Statistics for Runs A32a and I32\label{tab:statistics}}
    \tablehead{
    \colhead{} &
    \multicolumn{2}{c}{A32a} &
    \multicolumn{2}{c}{I32} \\
    \cline{2-5}
    \colhead{Quantity $f$} &
    \colhead{$\left<\left<f\right>\right> \times 10^2$} &
    \colhead{$\sigma_{\left<f\right>} \times 10^2$} &
    \colhead{$\left<\left<f\right>\right> \times 10^2$} &
    \colhead{$\sigma_{\left<f\right>} \times 10^2$}
    }
    \colnumbers
    \startdata
    $\rho v_x^2 / 2p_0$  & 0.27 & 0.081 & 0.35 & 0.112 \\
    $\rho v_y^2 / 2p_0$  & 0.15 & 0.054 & 0.22 & 0.095 \\
    $\rho v_z^2 / 2p_0$  & 0.17 & 0.044 & 0.19 & 0.066 \\
    $B_x^2 / 2p_0$       & 0.12 & 0.057 & 0.19 & 0.091 \\
    $B_y^2 / 2p_0$       & 1.18 & 0.576 & 1.51 & 0.823 \\
    $B_z^2 / 2p_0$       & 0.06 & 0.029 & 0.10 & 0.046 \\
    $\rho v_x v_y / p_0$ & 0.14 & 0.050 & 0.17 & 0.068 \\
    $-B_x B_y / p_0$     & 0.47 & 0.181 & 0.60 & 0.239 \\
    $p_{xy} / p_0$       & 0.20 & 0.058 & ---  & ---   \\
    $\left(p_{||}    - \left<p\right>_{t=0}\right) / p_0$ &
    -0.49 & 0.165 & --- & --- \\
    $\left(p_{\perp} - \left<p\right>_{t=0}\right) / p_0$ &
    0.15  & 0.083 & --- & --- \\
    \enddata
    \tablecomments{Single brackets denote spatial averages over the
    whole simulation domain; $\sigma_{X}$ is a standard deviation for a
    quantity $X$.}
   \end{deluxetable}

   \subsubsection{Volume-averaged stress}
\begin{figure}[!ht]
    \centering
    \includegraphics[width=\textwidth]{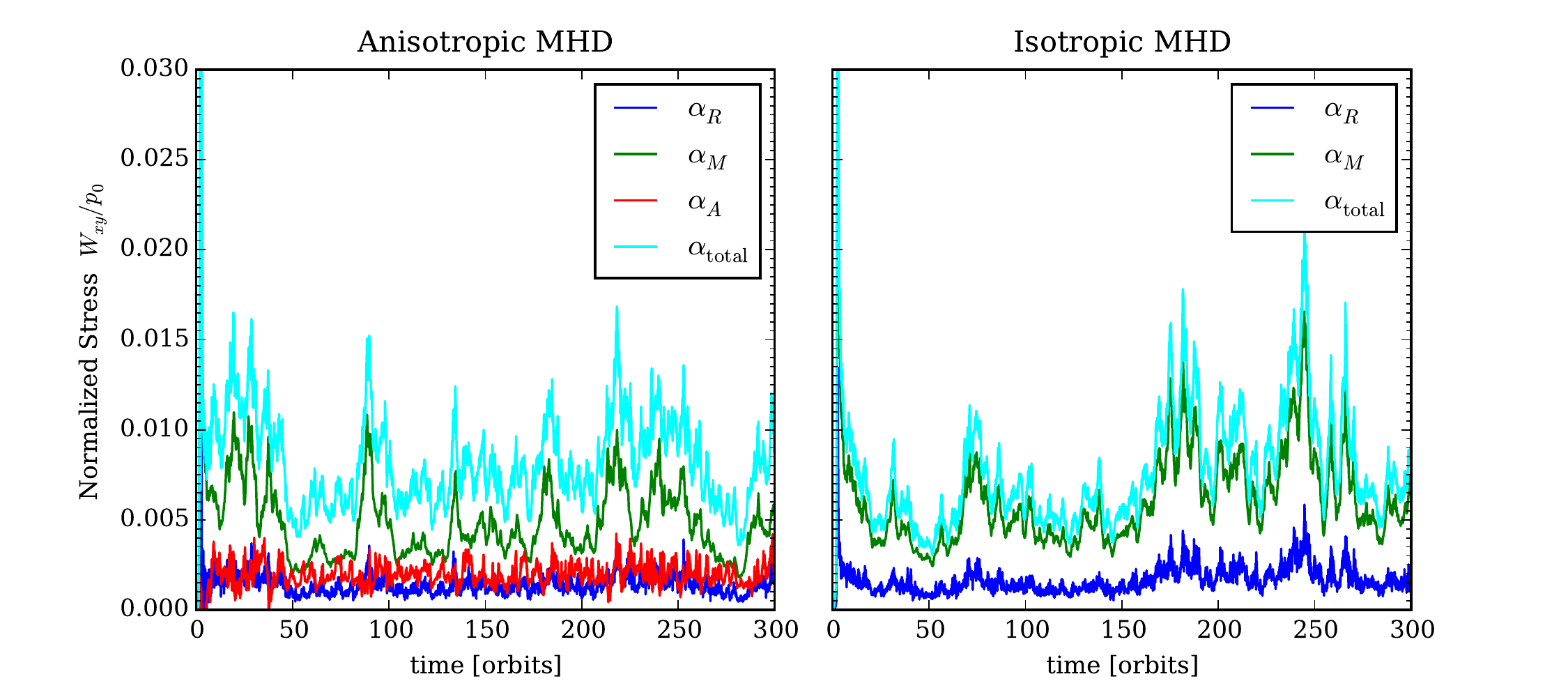}
    \caption{Time variation of $\alpha$-parameters averaged over the whole
    simulation domain in Runs A32a (left) and I32 (right).
    Contributions from the Reynolds (blue), Maxwell (green), anisotropic
    (red, only in the anisotropic run), and total (cyan) stresses are
    plotted with different colors.}
    \label{fig:stress-t}
   \end{figure}

   Figure~\ref{fig:stress-t} shows the time histories of the
   volume-averaged Reynolds, Maxwell, anisotropic, and total stresses
   for Runs A32a and I32.
   Each quantity shows a highly chaotic behavior.
   In particular, the Maxwell stress is strongly intensified
   intermittently with an interval that sometimes exceeds several dozen
   orbits.
   This is the reason why considerable durations such as 300 orbits are
   required to obtain a temporally averaged, representative value in
   shearing-box simulations of MRI-driven turbulence
   \citep{2003MNRAS.340..519W}.
   
\begin{figure}[!ht]
    \centering
    \includegraphics[width=0.6\textwidth]{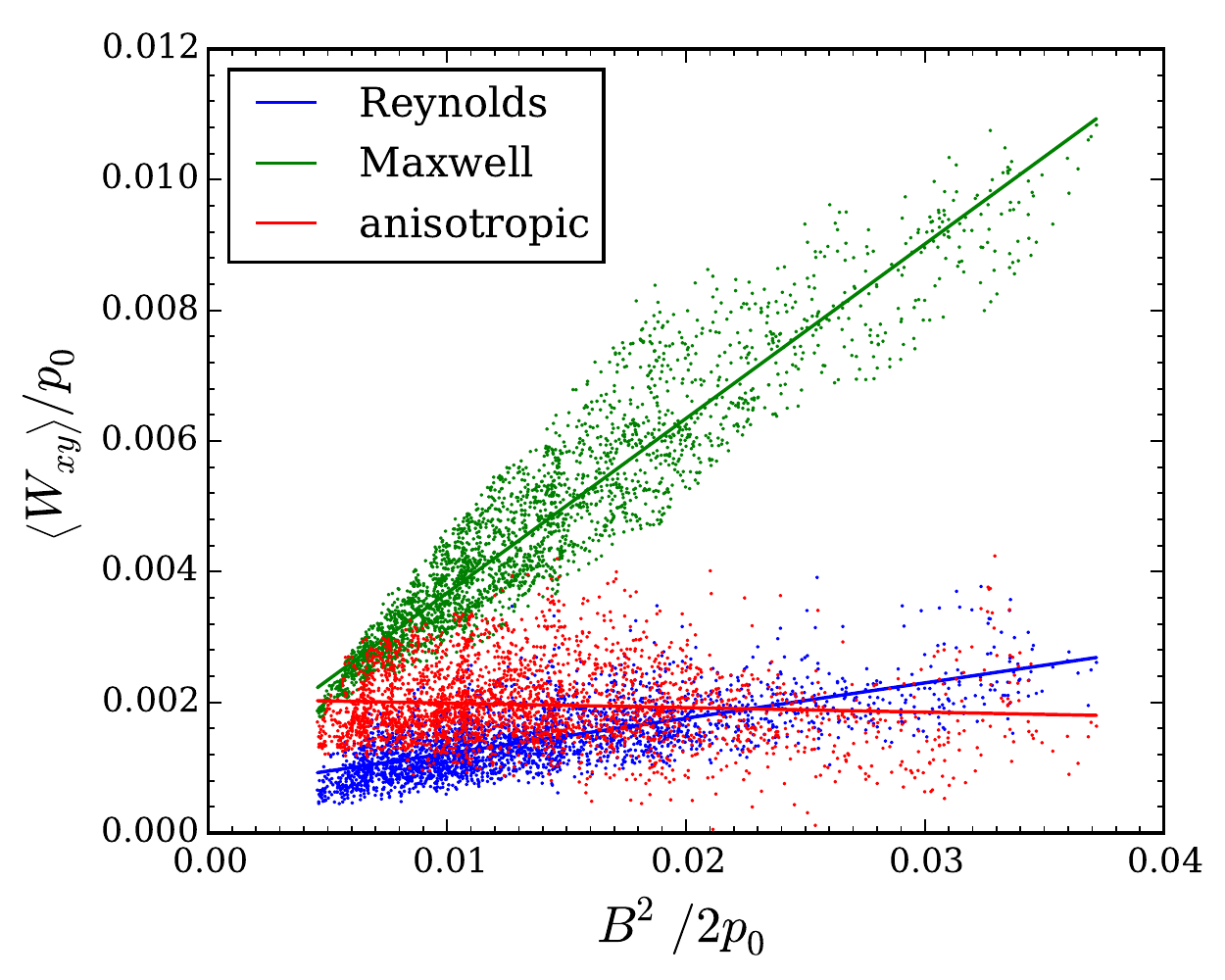}
    \caption{Dependence of each stress upon instantaneous magnetic
    energy.
    The solid lines are the results of linear fitting.}
    \label{fig:stress-b}
   \end{figure}

   Figure~\ref{fig:stress-b} also shows each stress averaged over the
   whole simulation domain after 50 orbits, but plotted as a function of
   the instantaneous magnetic energy.
   The green scattered points clearly indicate a strong correlation
   between the Maxwell stress and the magnetic energy.
   The slope of the regression line is 0.27, while the Reynolds stress
   shows weaker dependence with a slope of 0.054.
   The anisotropic stress, on the other hand, exhibits no clear
   dependence, with a slightly negative slope for the linear-regression
   line.
   This tendency looks quite different from that observed in an
   unstratified simulation (see Figure~3 in \cite{2006ApJ...637..952S}).
   For more details, we need to see the spatial structure, which will be
   discussed in the next section.
   
   \subsubsection{Vertical structure}
\begin{figure}[!ht]
    \centering
    \includegraphics[width=0.8\textwidth]{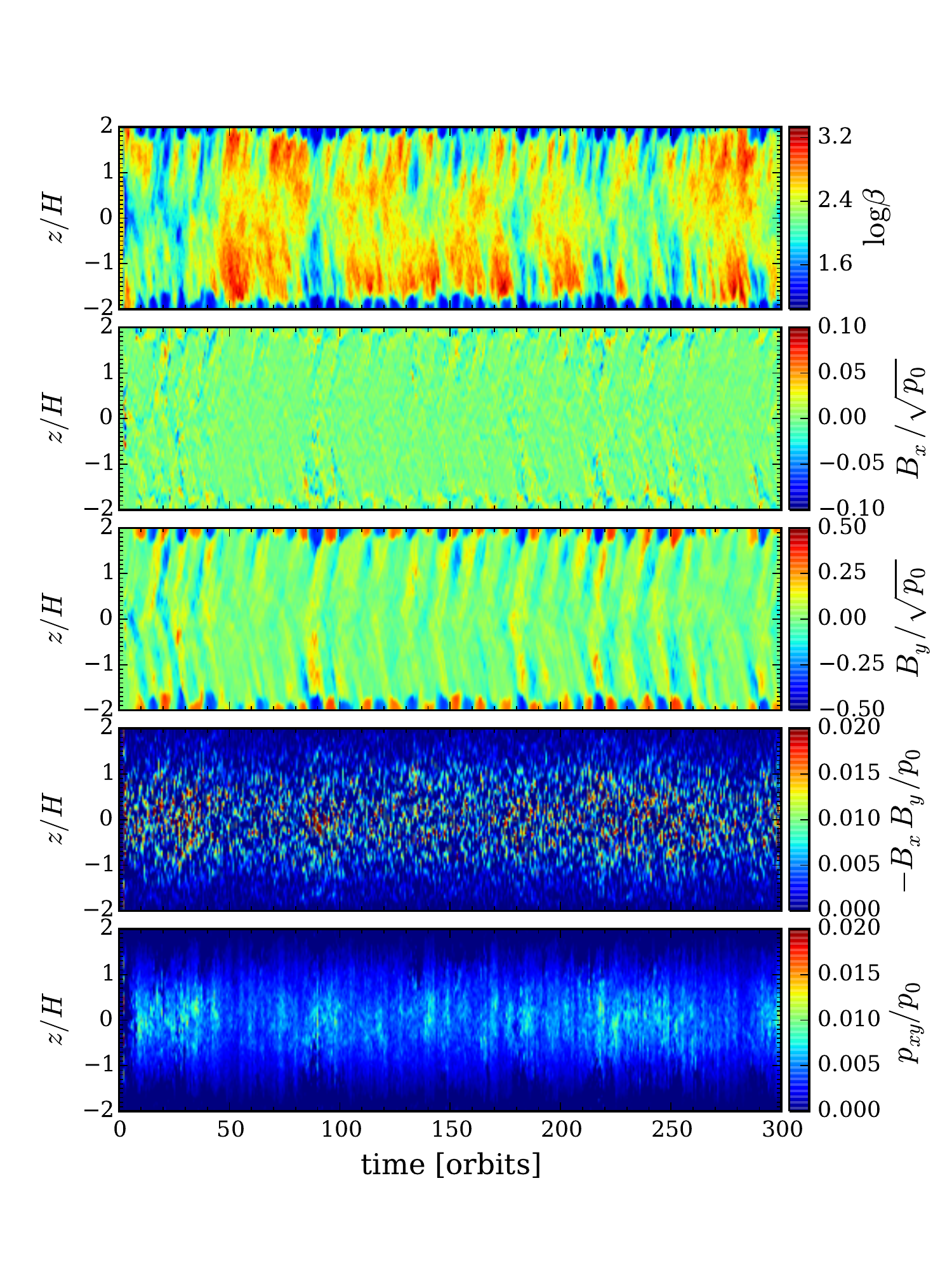}
    \caption{Horizontally averaged structure of various quantities in
    the anisotropic run as functions of vertical position and time in
    Run A32a.
    From top to bottom, we plot the plasma beta, the radial magnetic
    field, the azimuthal magnetic field, the Maxwell stress, and the
    anisotropic stress.}
    \label{fig:spacetime-a32}
   \end{figure}

\begin{figure}[!ht]
    \centering
    \includegraphics[width=0.8\textwidth]{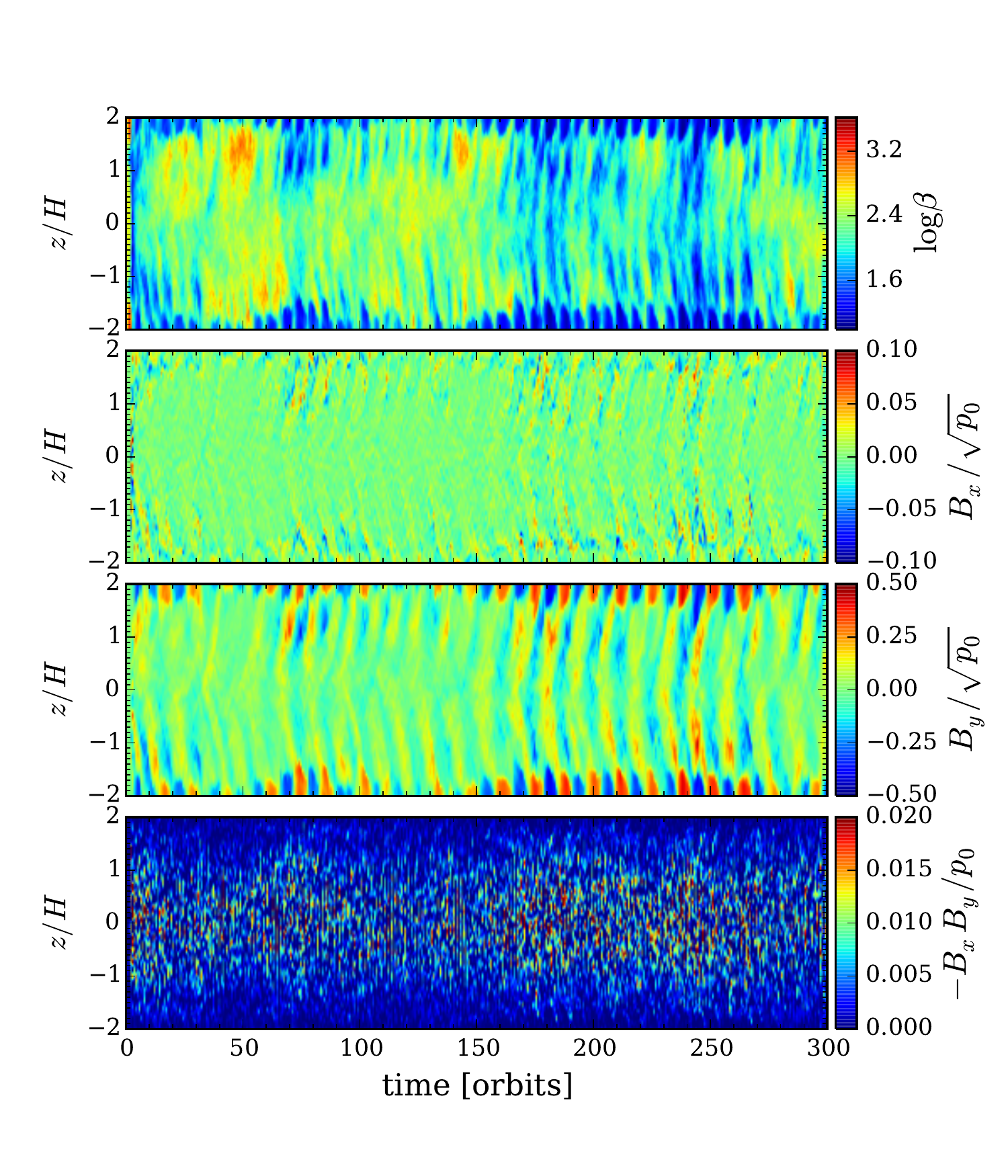}
    \caption{Space-time diagrams in isotropic Run I32 with the same
    format as in Figure~\ref{fig:spacetime-a32}.}
    \label{fig:spacetime-i32}
   \end{figure}

   Since the present system involves stratification, it becomes possible
   to study the vertical dependence of various quantities, which is a
   great advantage of the large-scale model.
   One graphical method that is helpful in understanding the vertical
   structure is a space-time diagram.
   It shows a quantity averaged along a horizontal plane as a
   two-dimensional function of both vertical position and time.
   Figures~\ref{fig:spacetime-a32} and \ref{fig:spacetime-i32} compare
   the space-time diagrams for Runs A32a and I32, respectively.
   From top to bottom, we plot the color contours of the plasma beta in
   logarithmic scale, the radial and azimuthal magnetic flux, the
   Maxwell stress, and the anisotropic stress (only in A32a), each of
   which is properly non-dimensionalized by the initial mid-plane
   pressure.
   
   From the top three panels, we can see that strongly magnetized
   patches generated near the disk mid-plane buoyantly rise toward the
   boundary regions.
   The magnetic flux piled up near the boundaries is apparently a
   spurious result of the use of periodic boundary conditions in the
   vertical direction.
   As we already mentioned, however, this structure does not affect the
   behavior inside the disk with $\left|z/H\right| \lesssim 1$, where
   the MRI is highly active.
   In particular, the bottom two panels show that neither the Maxwell
   nor the anisotropic stress has any corresponding structure near the
   boundaries, so the statistics of each stress in the previous section
   should not be significantly altered by the choice of the boundary
   condition.
   Looking at the sign of the azimuthal field in the rising patches (and
   also in the piled up field near the boundary), we can observe
   quasi-periodic reversals.
   This characteristic pattern is thought to be an indication of an
   underlying dynamo effect and has been observed commonly in previous
   MHD simulations in the collisional regime
   \citep[e.g.,][]{2007ApJ...668L..51P,2009ApJ...696.1021V,2010ApJ...713...52D,2010MNRAS.405...41G}.
   
\begin{figure}[!ht]
    \centering
    \includegraphics[width=\textwidth]{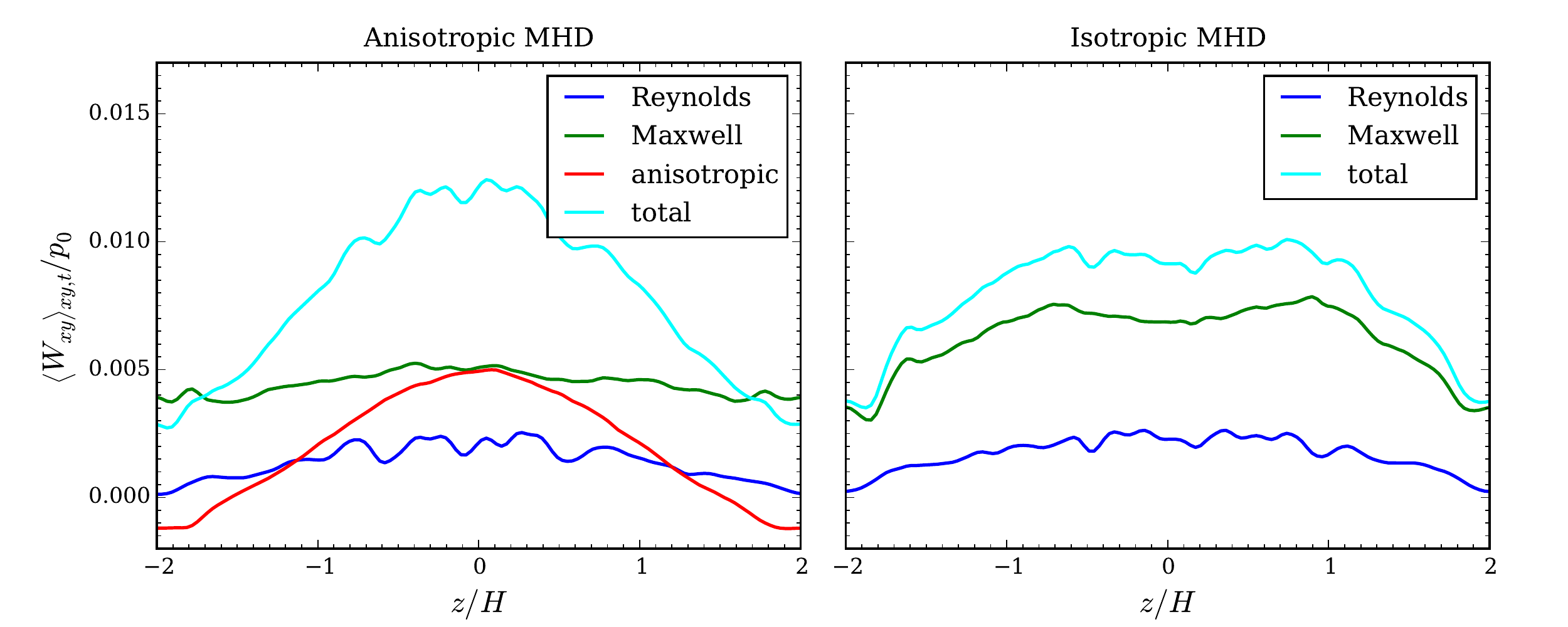}
    \caption{Comparison of the vertical structures of horizontally and
    temporally averaged stresses between Runs A32a and I32.} 
    \label{fig:stress-z}
   \end{figure}

   By comparison with Run I32 shown in Figure~\ref{fig:spacetime-i32},
   it can be said that the magnetic structure described here is
   qualitatively similar to that in the isotropic case.
   The vertical distribution of the anisotropic stress, however, is
   rather different from the other two components.
   To make this point clearer, we take a further average of time
   after 50 orbits, yielding the temporally and horizontally averaged,
   one-dimensional vertical profiles of the stress in
   Figure~\ref{fig:stress-z}, where each stress is colored in the same
   way as in Figure~\ref{fig:stress-t}.
   Again, the left and right panels represent the results for Runs A32a
   and I32, respectively.

   It is remarkable that the anisotropic stress localizes around the
   disk mid-plane, although the Maxwell and Reynolds components have
   broader structures over a wide range of height, just as in the
   isotropic MHD run.
   As a result, the total efficiency of angular-momentum transport also
   has a localized profile rather than a flat, or two-peak, distribution
   like that obtained in I32 and previous studies
   \citep{2010ApJ...713...52D}.
   This localization indicates that, for angular-momentum transport in
   collisionless disks, the effect of the background structure of an
   accretion disk could be more essential than expected in collisional
   disks.
   In particular, although unstratified-shearing-box simulations have
   predicted a magnitude of anisotropic stress comparable to that of the
   Maxwell stress, our stratified model shows that it is true only at
   the deep inside of the disk, $\left|z/H\right| \lesssim 0.5$, and
   that the anisotropic stress decreases gradually but more sharply than
   the other stresses as $z$ moves away from the disk mid-plane.
   Note that it is not surprising that the results of the unstratified
   models agree well with the activity around the mid-plane because the
   vertical gravity is proportional to $z$.
   
\begin{figure}[!ht]
    \centering
    \includegraphics[width=0.7\textwidth]{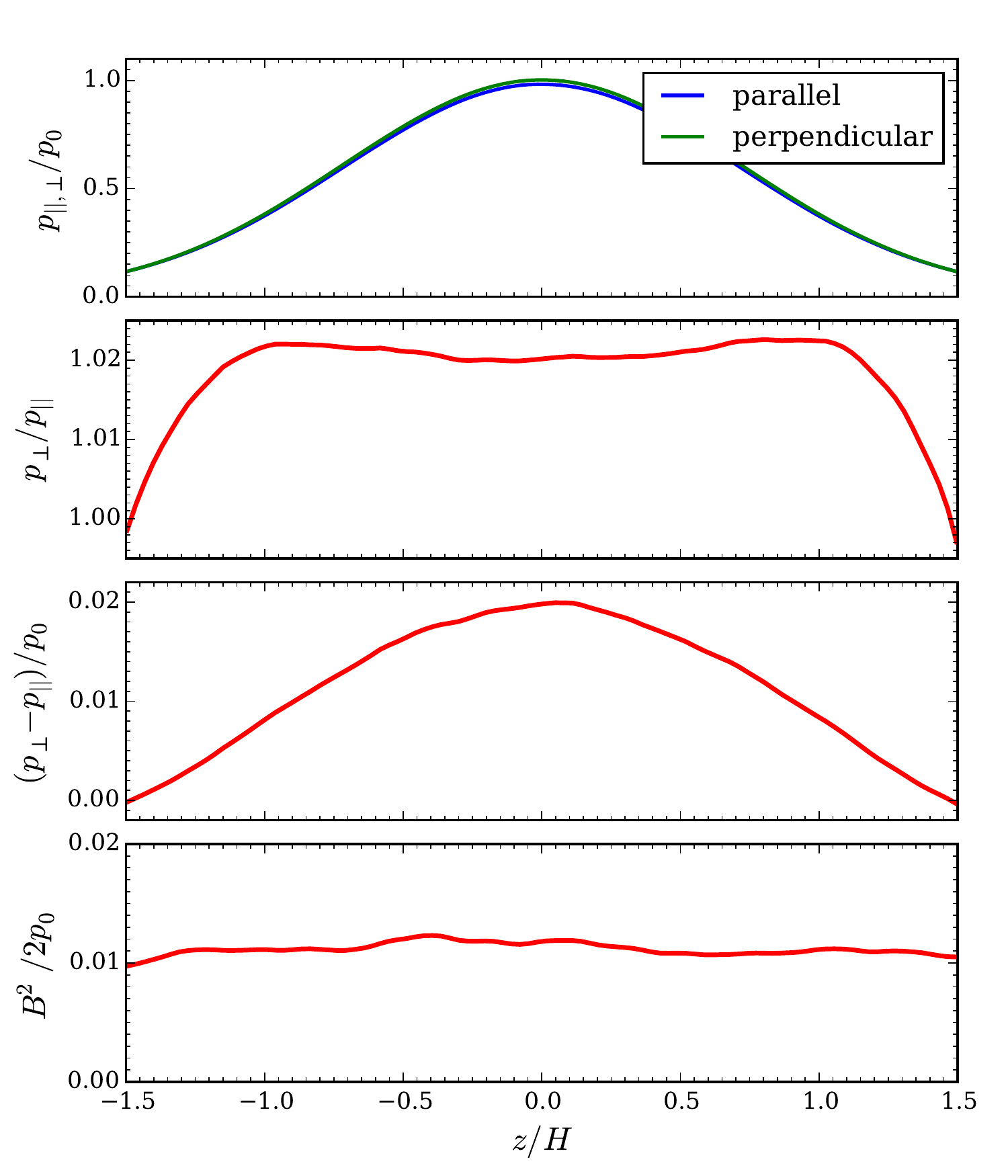}
    \caption{Horizontally and temporally averaged profiles of the
    variables related to anisotropic stress, $p_{xy}$.}
    \label{fig:p-z}
   \end{figure}

   The localization of the anisotropic stress seems to be explained as
   follows.
   $\alpha_A$, which is a part of $\alpha$ carried by the anisotropic
   stress, can be written as
   \begin{eqnarray}
    \alpha_{A}
     = \frac{p_{xy}}{p_0}
     \simeq \left(\frac{p_{\perp} - p_{||}}{B^2} \right)
     \left(\frac{-B_x B_y}{p_0}\right)
     = \left(
	\frac{\beta_{\perp} - \beta_{||}}{2}
       \right) \alpha_{M},
   \end{eqnarray}
   where $\alpha_M$ is the Maxwell stress normalized by $p_0$.
   This means that, even if anisotropy is somewhat uniform in the sense
   of $p_{\perp}/p_{||}$, $\alpha_A$ can vary largely depending on the
   magnitude of the diagonal components in the pressure tensor itself.
   This argument is illustrated in Figure~\ref{fig:p-z}.
   From top to bottom, plotted are the horizontally and temporally
   averaged profiles of the parallel and perpendicular pressure, the
   ratio and the difference between them, and the magnetic energy
   density, respectively, aer plotted within
   $\left|z/H\right| \leq 1.5$, where the spurious boundary effect is
   relatively weak.
   The first and second panels show that the anisotropy
   $p_{\perp}/p_{||}$ is rather uniform inside the disk,
   $\left|z/H\right| \simeq 1$, in the sense of their ratio, with the
   stratification maintained.
   The anisotropic stress, however, is determined by the difference of
   $p_{\perp}$ and $p_{||}$, which is plotted in the third panel, and it
   shows a convex profile due to the stratification.
   Since the magnetic energy density has a quite flat structure, as
   shown in the bottommost panel, the profile of anisotropic stress
   becomes nearly proportional to the stratified pressure.
   Although such a profile of the anisotropic stress might seem an
   apparent consequence, this is undoubtedly the first quantitative
   estimate of the anisotropic stress in a collisionless accretion disk
   including the disk scale.
   
  \subsection{Resolution dependence}
  \label{subsec:resolution}
  Convergence of solutions with respect to the number of grid points is
  an issue of importance for numerical simulations.
  The grid convergence of statistics in a shearing-box model during a
  saturated stage of MRI has also been investigated intensively.
  Focusing on calculations with no net magnetic flux, it is known that
  the turbulent stress in cases without explicit dissipation and
  vertical gravity becomes smaller and smaller without bound as a
  higher resolution is employed
  \citep{2007ApJ...668L..51P,2007A&A...476.1113F,2009ApJ...694.1010G}.
  This absence of convergence has been a long-standing mystery of
  shearing-box models.
  A resolution study using the Athena code reported in
  \cite{2010ApJ...713...52D}, however, claimed that they observed
  convergence of the saturation amplitude of MRI-driven turbulence for
  resolutions up to 128 grid points per scale height, once the
  vertical gravity and the resultant stratification were retained. 
  In contrast, \cite{2014ApJ...787L..13B} revisited the same problem
  with ampler computational resources using the PLUTO code, and observed
  {\it no} evidence of convergence with resolutions up to 200 grid
  points per scale height.
  The grid convergence of turbulent stress in a shearing-box model is
  still an open question.

  In the case of our model as well, it is significant to assess the
  dependence of the transport efficiency upon the number of grid
  points.
  We list in Table~\ref{tab:runs} the results using different
  resolutions with 32 and 64 grid points per scale height in Runs A32a
  and A64 (I32 and I64 for isotropic pressure).
  Note that the Reynolds and Maxwell stresses in Run I64 agree
  remarkably well with those obtained in Run S64R1Z4 reported in
  \cite{2010ApJ...713...52D}, which corresponds to the same calculation
  as in our Run I64.
  Run A64, on the other hand, shows a more sensitive increase in every
  component of the stress, and the total transport efficiency exceeds
  the isotropic case.
  The duration of the high-resolution run, however, is limited to only
  20 orbits owing to the restriction of currently available
  computational resources.
  This duration does not seem sufficient at all for obtaining a
  meaningful representative value from highly chaotic turbulence, since
  the interval of intermittent behavior can reach several dozen orbits,
  as we already mentioned.
  We leave a more precise and robust study on convergence in simulations
  lasting several hundreds of orbits for future work. 

  \subsection{Dependence on gyrotropization model}
  \label{subsec:gyro}
  To deal with an anisotropic pressure tensor, our model introduces a 
  timescale or a frequency to control how fast the pressure tensor
  approaches its gyrotropic asymptote. 
  It is necessary to clarify the dependence of our results upon this
  parameter.
  The gyrotropization frequency enters into the system through the last
  term in Equation (\ref{eq:eq_energy}), where 
  \begin{eqnarray}
   \nu_g = \left(\frac{\left|\mathbf{B}\right|}{B_0}\right)\nu_{g0},
  \end{eqnarray}
  is set to be proportional to the local magnetic-field strength
  normalized by a fiducial value, $B_0$.
  In all of our calculations, $B_0$ is fixed at the initial maximum
  value of the magnetic field for $\beta=100$, i.e.,
  $B_0 = \sqrt{2p_0/10^{2}}$, which is also a typical field strength
  during the saturated stage.
  Runs A32a, A32d, and A32e in Table~\ref{tab:runs} employ different
  values of $\nu_{g0}$, namely, $10^{13}$, $10^{8}$ and $10^{3}$ in
  units of the rotation frequency of the disk, $\Omega$.
  These choices of $\nu_{g0}$ roughly yield
  $\nu_{g0} \Delta t \simeq 10^{10}$, $10^{5}$, and $1$, respectively,
  at a site of a typical magnetic field,
  using a simulation time step defined to satisfy the CFL condition.
  We therefore expect quantitatively similar results for Runs A32a
  and A32d, where $\nu_{g}\Delta t \gg 1$, and the statistics shown in
  Table~\ref{tab:runs} successfully demonstrate it.

\begin{figure}[!ht]
   \centering
   \includegraphics[width=\textwidth]{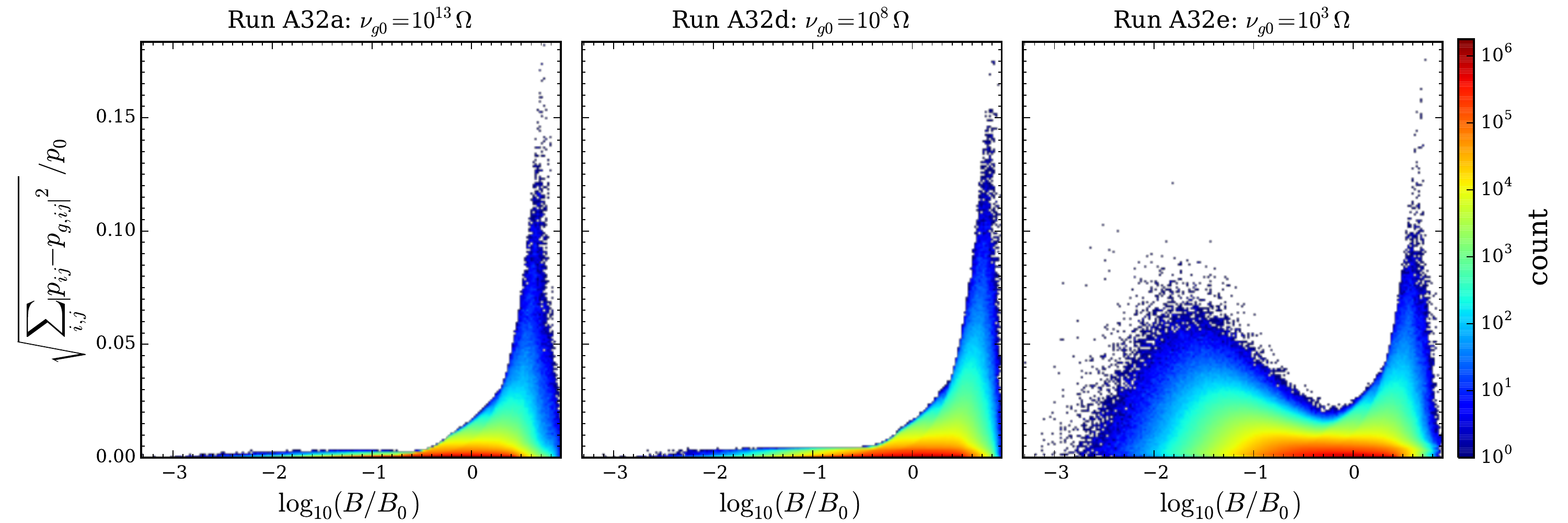}
   \caption{Occurrence frequencies of a measure of non-gyrotropy and
   magnetic-field strength, for Runs A32a, A32d, and A32e using different
   gyrotropization frequencies.}
   \label{fig:nongyro}
  \end{figure}

  Run A32e, on the other hand, shows that the Maxwell stress increased
  by $\sim 40 \%$ whereas the Reynolds and anisotropic contributions
  remained unchanged.
  The magnetic energy is also increased by 35\%.
  This enhancement may be traced to the reduction of released magnetic
  energy via magnetic reconnection under the presence of large
  non-gyrotropy.
  In HHA16, it was illustrated that eigenmodes that are not present
  in a gyrotropic framework can support a current sheet without
  resulting in an explosive reconnection event and releasing
  magnetically stored energy.
  To support this assertion, we present plots of the occurrence
  frequencies of non-gyrotropy in Figure~\ref{fig:nongyro}.
  The vertical axis indicates a measure of non-gyrotropy, which is
  defined as an $L_2$-norm of a residue of a pressure tensor from its
  gyrotropic limit, as normalized by the mid-plane pressure;
  \begin{eqnarray}
   \Delta\hat{p}
    &\equiv&
    \frac{1}{p_0}\sqrt{\sum_{i,j}\bigl|p_{ij}-p_{g,ij}\bigr|^2},
    \nonumber \\
    &=&
    \frac{1}{p_0}
    \sqrt{
    (p_{xx}-p_{g,xx})^2+(p_{xy}-p_{g,xy})^2+\dots+(p_{zz}-p_{g,zz})^2
    }.
    \nonumber
  \end{eqnarray}
  The horizontal axis shows the magnetic-field strength on a logarithmic
  scale, and the number of cells counted during 50--100 orbits is shown
  by the color contour.
  From left to right, the panels show the results for Runs A32a, A32d,
  and A32e, respectively.

  Again, the two left panels for Runs A32a and A32d show the quite
  similar results; a majority of data crowds around a narrow region
  along $\Delta\hat{p} \simeq 0$.
  Note that $\nu_g \Delta t \gg 1$ holds over the entire range field
  strengths in these two runs.
  In the rightmost panel for the smallest value of $\nu_{g0}$, however,
  a large non-gyrotropy with at most $\Delta\hat{p} \simeq 0.1$
  remains in the weakly magnetized regions, which can be thought as sites
  where magnetic reconnection takes place.
  The extent to which this highly non-uniform non-gyrotropy alters the
  dynamics of reconnection is still unclear.
  Nevertheless, in a qualitative sense at least, our result seems
  consistent with the suppression of reconnection by the non-gyrotropic
  entropy modes as pointed out in HHA16.
  
  The enhanced stress in Run A32e highlights the important role of
  deviation from gyrotropic pressure, which updates the previous result
  of the shearing-box simulation with the gyrotropic framework by
  \cite{2006ApJ...637..952S}.
  Even though the volume occupancy of weakly magnetized regions with
  finite non-gyrotropy is not very large, it can have a significant
  impact upon the full large-scale dynamics through the process of
  magnetic reconnection.
  This is a good example of the multi-scale coupling typical in
  collisionless plasmas, and we have suggested a new aspect of
  micro-physics to be taken into account in large-scale collisionless
  accretion disks.

   \subsubsection{Combined model of rotation and gyrotropization}
   \label{subsec:rotation}
   For the purpose of showing one of the possible directions for
   improvement of the gyrotropization model, we have attempted numerical
   experiments to include the effect of rotation of a pressure tensor in
   addition to gyrotropization;
   we solve
   \begin{eqnarray*}
    \frac{\partial p_{ij}}{\partial t}
     = \Omega_{c}
     \left(
      \varepsilon_{ikl} p_{jk} \hat{b}_l +
      \varepsilon_{jkl} p_{ik} \hat{b}_l
	   \right)
     - \nu_g
     \left(
      p_{ij} - p_{g,ij}
     \right),
   \end{eqnarray*}
   instead of only the gyrotropization effect, whose right-hand side
   includes only the last term with $\nu_g$.
   This is motivated by the fact that the original form of this operator
   simply describes the rotation of a pressure tensor by cyclotron
   motion of particles (see Equation (12) in HHA16).
   By solving this term analytically in an operator-splitting manner, the 
   effect that a particle incoming into a neutral sheet quickly changes
   its direction of cyclotron motion, which is one of the essential
   phenomena in magnetic reconnection, may be incorporated into the
   system.
   In practice, since our model employs a one-fluid framework, we have
   to determine the direction in which particles rotate.
   In the present application, suppose that the pressure is mainly
   supported by ions and a positive value is adopted for $\Omega_c$,
   which is defined in the same way as $\nu_g$;
   \begin{eqnarray*}
    \Omega_c = \left(\frac{\left|\mathbf{B}\right|}{B_0}\right)\Omega_{c0}.
   \end{eqnarray*}

   The role of this rotation term, particularly on magnetic
   reconnection, is discussed in Appendix~\ref{app:rotation} using a
   one-dimensional Riemann problem that imitates the self-similar
   Petschek-type reconnection layer, and we find that the fast rotation
   of a pressure tensor around a local magnetic-field line behaves in a
   qualitatively similar way to the previous model, whereas
   gyrotropization is not guaranteed explicitly.
   In spite of the similarity, inclusion of this original rotation term
   still has an advantage.
   We observed that, when applied to stratified-disk simulations, it
   becomes possible to employ smaller $\nu_{g0}$ than listed in
   Table~\ref{tab:runs} for the first time by using the combined model;
   otherwise, the simulation with, for example, $\nu_{g0}=10\Omega$
   breaks down owing to the appearance of negative density.
   On the contrary, a merit of incorporating with the gyrotropization
   model is to guarantee consistency with the gyrotropic limit in an
   asymptotic manner by reducing non-gyrotropy explicitly.

   \subsubsection{Application to stratified-disk simulations}
   In this section, let us apply the combined model introduced above to
   our stratified-disk model.
   For simplicity, we assume $\Omega_{c0} = \nu_{g0}$ here.
   The simulation parameters and the resultant stress averaged both in
   space and time are summarized in Table~\ref{tab:runs2} with the
   same format as in Table~\ref{tab:runs}.
   Each run is labeled with R (with rotation).
   From Run R32a to Run R32d, the frequency common to the
   gyrotropization and rotation procedures is changed.
   
   \begin{deluxetable}{cccccccc}
    \tablecaption{Simulation summary for the rotation + gyrotropization
    model\label{tab:runs2}}
    \tablehead{
    \colhead{Run} &
    \colhead{Resolution} &
    \colhead{Orbits} &
    \colhead{$\beta$} &
    \colhead{$\nu_{g0}/\Omega$} &
    \colhead{$\left<\left<\rho v_x v_y\right>\right>/p_0$} &
    \colhead{$\left<\left<-B_x B_y\right>\right>/p_0$} &
    \colhead{$\left<\left<p_{xy}\right>\right>/p_0$}
    }
    \colnumbers
    \startdata
    R32a & $32/H$ & 300 & $10^2$ & $10^{13}$  & 0.0014 & 0.0045 & 0.0019 \\
    R32b & $32/H$ & 300 & $10^2$ & $10^{ 8}$  & 0.0015 & 0.0048 & 0.0020 \\
    R32c & $32/H$ & 300 & $10^2$ & $10^{ 3}$  & 0.0016 & 0.0057 & 0.0021 \\
    R32d & $32/H$ & 300 & $10^2$ & $10^{ 1}$  & 0.0019 & 0.0121 & 0.0090 \\
    \enddata
   \end{deluxetable}

\begin{figure}[!ht]
    \centering
    \includegraphics[width=0.85\textwidth]{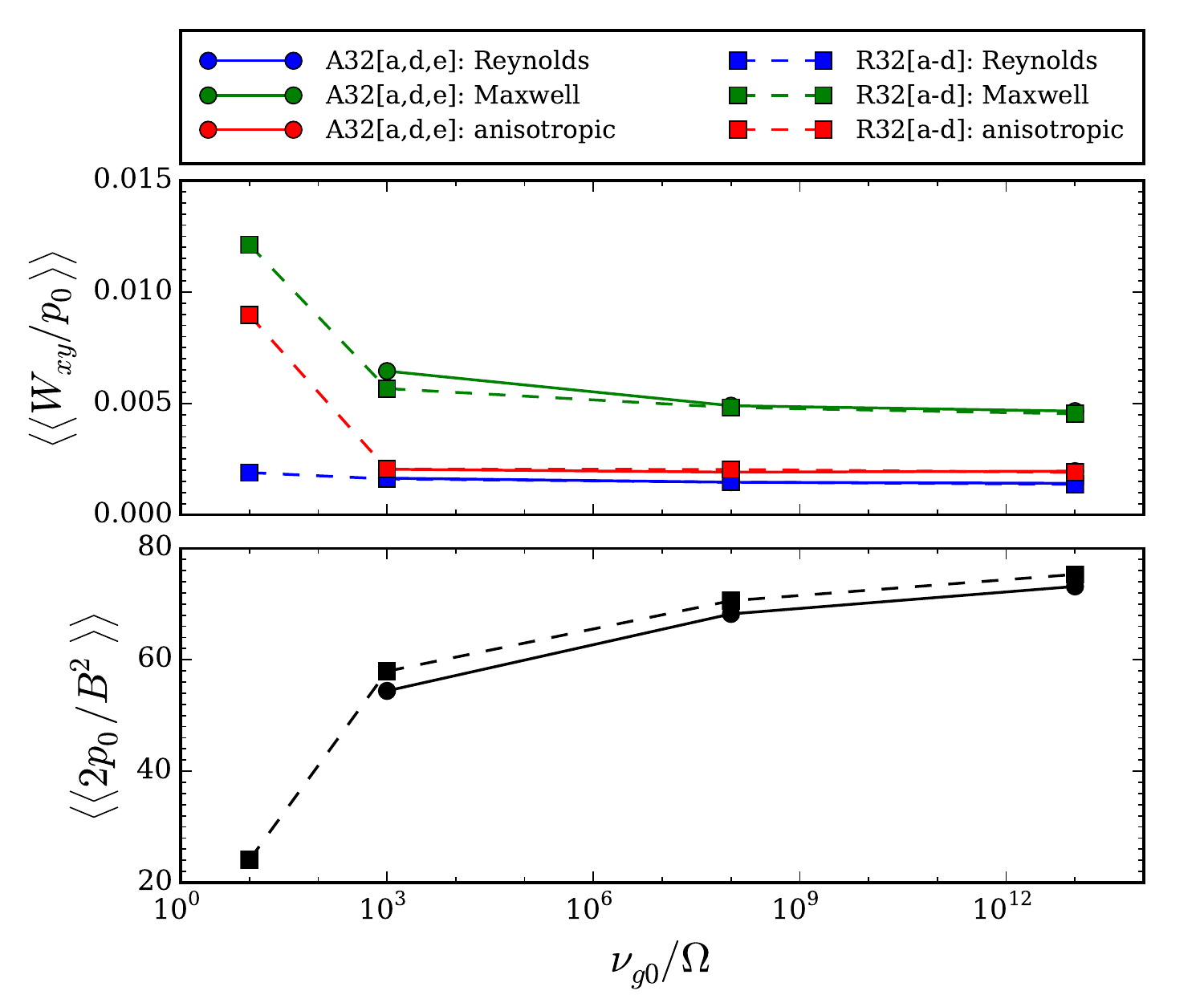}
    \caption{Dependence of stress and plasma beta upon the
    gyrotropization and cyclotron frequency.
    The circles connected with solid lines are the cases with only
    gyrotropization, and the squares connected with dashed lines
    indicate the result of the gyrotropization and rotation model.}
    \label{fig:gyro-rotate}
   \end{figure}

\begin{figure}[!ht]
    \centering
    \includegraphics[width=0.85\textwidth]{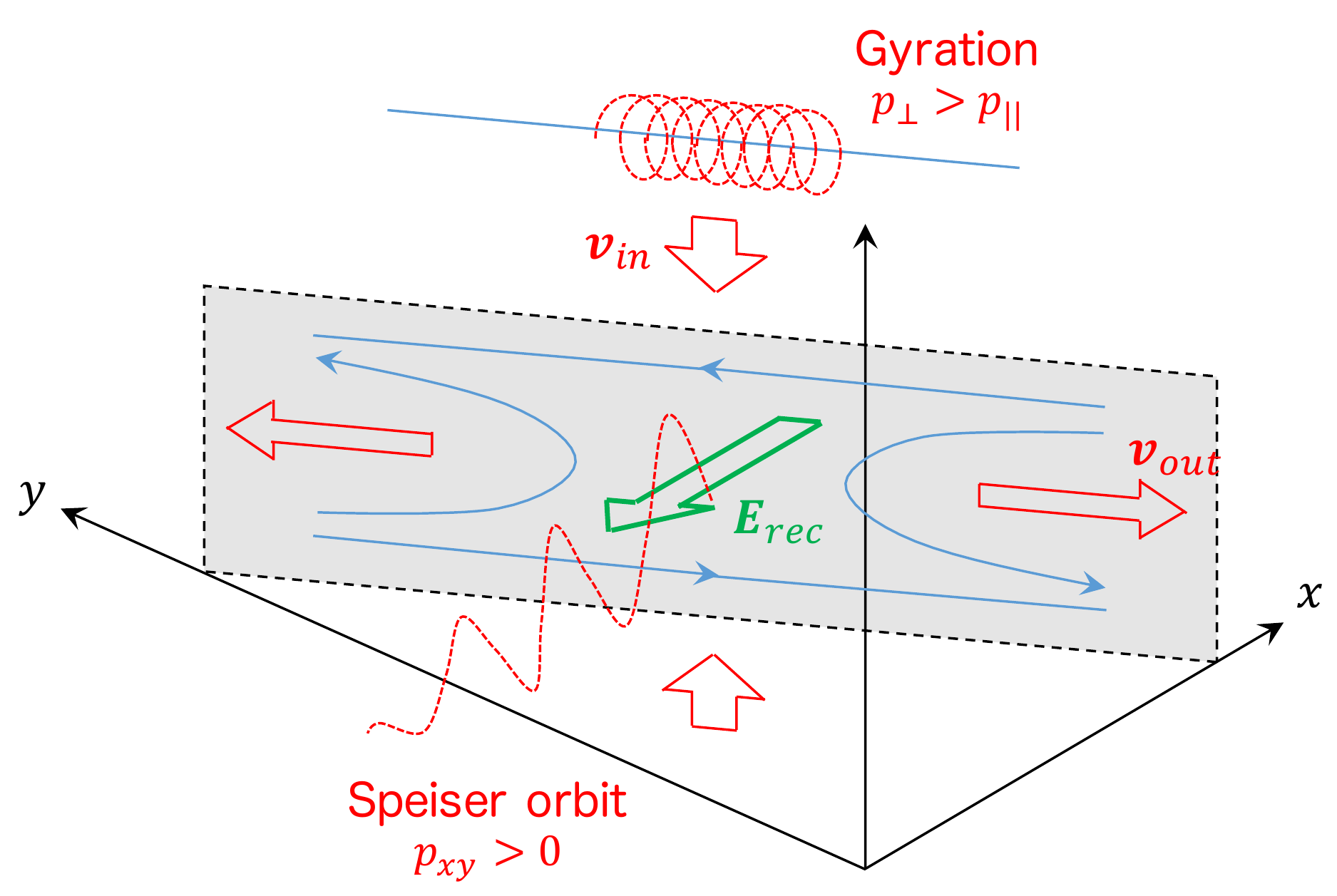}
    \caption{A schematic diagram of a reconnection layer expected in a
    shearing-box simulation.}
    \label{fig:speiser}
   \end{figure}

   The dependence of the stress on $\nu_{g0}$ is illustrated in
   Figure~\ref{fig:gyro-rotate} along with the averaged plasma beta.
   The A32 and R32 series are represented by circles and squares,
   respectively.
   We can clearly see that both series are in qualitative and
   quantitative agreement for $\nu_{g0} \gtrsim 10^3 \Omega$, which
   ensures that the current model is an appropriate extension of the
   original model when a pressure tensor is well gyrotropized.
   The intensive enhancement in the Maxwell and anisotropic stresses is,
   however, observed when $\nu_{g0}$ is further decreased to $10\Omega$
   in Run R32d, and the magnetic energy is also roughly tripled compared
   with the case of $\nu_{g0}=10^3\Omega$.
   Although these facts seem to imply a more efficient {\it dynamical}
   suppression of magnetic reconnection by a non-gyrotropic pressure,
   the increase not only in the Maxwell stress but also in the
   anisotropic stress may be related to a process slightly different from
   the case in A32e or R32c.
   For example, it is known in the kinetic description that charged
   particles entering a reconnection layer experience Speiser orbits
   toward the direction of the electric field along an X-line.
   The particular situation in a shearing box is illustrated in
   Figure~\ref{fig:speiser}.
   Since the magnetic field is continuously stretched in the
   $x$-direction by the MRI and in the $y$-direction by background
   differential rotation, reconnection is expected to occur in a plane
   inclined from both the $\left(x,z\right)$- and
   $\left(y,z\right)$-planes, like the shaded region.
   Thus, the population in Speiser orbits tends to have a correlation of
   $\left<v_x v_y\right>>0$, which contributes to positive $p_{xy}$
   in quite non-gyrotropic manner.
   Getting back to our fluid model, although each particle's orbit is
   not resolved, it may be possible to generate this non-gyrotropic
   $p_{xy}$ by distorting and rotating the pressure of incoming flux
   with large $p_{\perp}$.
   This effect cannot be captured when only gyrotropization is assumed,
   where the anisotropy around the magnetic field remains in a constant
   direction.
   Demonstrating this hypothesis, however, requires a more detailed
   study of magnetic reconnection itself in an isolated system.
   In any case, the result here emphasizes the striking dependence of
   the angular-momentum-transport efficiency upon the ratio between the
   cyclotron frequency and the disk's rotation frequency.
   Note that, since the previous PIC simulations by
   \cite{2015PhRvL.114f1101H} assumed a small ratio,
   $\Omega_c/\Omega=10$, to save computational resources, it may be
   possible to overestimate the suppression effect due to the
   non-gyrotropic distribution function.
   It is to be hoped that future research will clarify this point.
   
 \section{Discussion and summary}
 \label{sec:summary}
 In this paper, we have conducted a series of stratified-shearing-box
 simulations in the collisionless regime using the recently developed
 kinetic-MHD model, which can handle finite deviation from the
 gyrotropic pressure.
 This is the first approach to the large-scale dynamics of collisionless
 accretion disks.
 In particular, it is designed to bridge the gap between
 double-adiabatic simulations assuming gyrotropic pressure and fully
 kinetic simulations, which cannot handle disk scales.

 The main results of importance are summarized as follows:
 \begin{enumerate}
  \item The volume- and time-averaged total efficiency of
	angular-momentum transport remains at the same level as in
	isotropic MHD simulations for large gyrotropization rates.
  \item The distribution ratio of the Reynolds, Maxwell, and anisotropic
	stresses agrees with unstratified simulations only near the
	disk's mid-plane, and anisotropic stress decreases outwardly.
  \item The results are not affected by the choice of the artificial
	parameter used to determine the gyrotropization rate, as long as
	it is sufficiently large.
  \item Once the gyrotropization rate approaches the dynamical timescale
	of the disk, finite non-gyrotropy around neutral sheets tends to
	enhance the Maxwell stress.
 \end{enumerate}
 Although the result 1 looks the same as the argument claimed in a
 previous study of unstratified simulations using the double-adiabatic
 model \citep{2006ApJ...637..952S}, the contribution from the
 anisotropic stress is only comparable to the Reynolds stress, rather
 than to the Maxwell stress as predicted in the unstratified case.
 The properties of the unstratified-shearing-box model are well
 reproduced only around the disk's mid-plane, where the effect of
 vertical gravity is weak and stratification can be ignored to a good
 approximation.
 The stratified pressure, however, yields the strong dependence of the
 anisotropic stress upon the vertical position.
 Specifically, as stated in result 2, the anisotropic stress decreases
 outwardly away from the mid-plane, reflecting stratification of the
 diagonal pressure, whereas the Maxwell and Reynolds stresses have
 broader distributions.
 This fact apparently suggests that the large-scale structure is more
 essential for angular-momentum transport in a collisionless accretion
 disk than expected in a collisional disk where thermal pressure cannot
 carry angular momentum.
 Our results represent an important foothold for a deeper understanding
 of the global behavior of collisionless disks in the future.

 A self-similar solution for Sgr A* based on the ADAF model
 \citep{1994ApJ...428L..13N} predicts that the ratio of the cyclotron
 frequency to the rotational frequency of the disk largely depends upon
 an orbital radius, but is, in general, expected to be much greater than
 unity.
 Result 3 relieves us of a distress about a practical choice for a
 gyrotropization frequency.
 When it is artificially set to a value comparable to a dynamical
 frequency, on the other hand, the energy released through magnetic
 reconnection could be suppressed by finite non-gyrotropy, and an
 efficient transport is likely to be sustained by the increased Maxwell
 stress.
 Result 4 is qualitatively consistent with PIC simulations in
 \cite{2015PhRvL.114f1101H}, though we must note that the physics
 suppressing the reconnection process would be quite different.
 It is a highly challenging matter of interest to improve our current 
 model to capture this suppression effect, which is believed to reflect
 {\it micro}-physics in part, and a large gyrotropization rate is
 maintained outside of the current sheets, where {\it macro}-physics
 plays a role.
 Note that the role of non-gyrotropy in the context of MRIs was not
 discussed in fully kinetic PIC simulations by
 \cite{2015PhRvL.114f1101H}, where suppression of magnetic reconnection
 and the resultant enhancement of angular-momentum transport were
 accounted for only in terms of gyrotropic anisotropy with
 $p_{||}>p_{\perp}$.
 Although there have been theoretical and numerical studies on
 non-gyrotropic electron pressure, which strongly affects the physics
 of resistivity, and hence magnetic reconnection, through the
 generalized Ohm's law
 \citep[e.g.,][]{1993GeoRL..20.1207H,1994JGR....9911177H,2003PhPl...10.1595Y},
 its dynamical role has not been discussed with interest.
 The simple test problem of mimicking a one-dimensional reconnection
 layer and its indirect application to a larger system demonstrated here
 emphasizes the necessity of seeing the importance of the non-gyrotropic
 components of a pressure tensor in a new light.
 
 It should be noted that our model currently does not include the effect
 of resistivity, which is another large issue for determining the
 saturation amplitudes of MRI-driven turbulence.
 The lack of resistivity is due to the distribution of thermal energy
 generated by Joule heating of each component of a pressure tensor that
 cannot be determined within the fluid framework.
 This means that magnetic reconnection occurs only through numerical
 dissipation in the current model, and the heating rate of each
 component of a pressure tensor is not under control.
 We, however, could employ a model of the heating rates, for example,
 based on the idea that the Joule heating is mainly carried by
 electrons that isotropize instantaneously.
 Conversely, it is possible to construct a model of anisotropic heating
 of ions.
 Investigating the dependence of the present results upon resistivity
 models is also left as future work to be discussed.

 \acknowledgements
 We would like to thank James M. Stone for the thoughtful comments on an
 early stage of this work.
 We would also like to thank Takanobu Amano for the useful and
 encouraging discussions.
 Numerical computations were carried out in part on Cray XC30 at the
 Center for Computational Astrophysics, National Astronomical
 Observatory of Japan.
 This work was supported by JSPS KAKENHI Grant Number 26$\cdot$394.

 \appendix
 
 \section{Orbital-Advection Scheme with Anisotropic Pressure}
 \label{app:advection} 
 The orbital-advection scheme introduced in \cite{2010ApJS..189..142S} is
 a numerical technique to integrate MHD equations in a shearing box
 accurately and efficiently by decomposing the system into two
 subsystems;
 one is a simple linear system which describes advection
 by background-differential rotation, and the other is a standard
 hyperbolic system with modified shearing-source terms.
 Since a detailed numerical implementation is provided in
 \cite{2010ApJS..189..142S}, here, we only reproduce the practical
 equations employed in our simulation code for the sake of completeness,
 along with comments on a few modifications required to apply the
 orbital-advection scheme to the kinetic-MHD model with an
 anisotropic-pressure tensor.
 
 Let us decompose Equations (\ref{eq:eq_mass})--(\ref{eq:eq_energy}).
 The subsystem for linear advection can be written simply as follows:
 \begin{eqnarray}
  \frac{\partial \rho}{\partial t}
  + v_K \frac{\partial \rho}{\partial y} = 0,
  \label{eqC:linear_mass} \\
  \frac{\partial \rho \mathbf{v}^\prime}{\partial t}
  + v_K \frac{\partial \rho \mathbf{v}^\prime}{\partial y} = 0, \\
  \frac{\partial \mathbf{B}}{\partial t}
  - \nabla \times
  \left(v_K \hat{\mathbf{e}}_y \times \mathbf{B}\right) = 0, \\
  \frac{\partial \mathbf{E}^\prime}{\partial t}
  + v_K \frac{\partial \mathbf{E}^\prime}{\partial y}
  = \mathbf{s}_{E^\prime}, \label{eqC:linear_energy}
 \end{eqnarray}
 where $v_K=-q\Omega x$ represents the background Keplerian velocity,
 $\mathbf{v}^\prime=\mathbf{v}-v_K\hat{\mathbf{e}}_y$ is the deviation
 from the Keplerian rotation,
 $\mathbf{E}^\prime=\rho\mathbf{v}^\prime\mathbf{v}^\prime+\mathbf{p}+\mathbf{BB}$
 is a generalized energy tensor, and
 \begin{eqnarray}
  \mathbf{s}_{E^\prime} =
   - \left(\rho v_x^\prime v_y^\prime + p_{xy} - B_x B_y\right)
   \frac{dv_K}{dx}
   \left(
    \begin{array}{ccc}
     0 \ \ \ & 1 \ \ \ & 0 \ \ \ \\
     1 \ \ \ & 2 \ \ \ & 1 \ \ \ \\
     0 \ \ \ & 1 \ \ \ & 0 \ \ \ \\
    \end{array}
   \right).
 \end{eqnarray}
 Since the advection velocity is purely along the $y$-axis and constant
 in time, these equations can be integrated using their analytical
 solutions without constraint upon the CFL condition.
 In practice, this procedure is implemented in a similar way to the
 shearing periodic boundary condition, although we must take special care
 of the divergence-free condition for a magnetic field.
 The right-hand side of Equation (\ref{eqC:linear_energy}) cannot be
 gathered up in a flux form and is treated as a source term separately.

 By subtracting Equations
 (\ref{eqC:linear_mass})--(\ref{eqC:linear_energy}) from Equations
 (\ref{eq:eq_mass})--(\ref{eq:eq_energy}), we find that the
 remaining part yields the following subsystem:
 \begin{eqnarray}
  \frac{\partial \rho}{\partial t}
  + \nabla \cdot \left( \rho \mathbf{v}^\prime \right) = 0,
  \label{eqC:mass} \\
  \frac{\partial \rho\mathbf{v}^\prime}{\partial t}
  + \nabla \cdot
  \left(
  \rho\mathbf{v}^\prime\mathbf{v}^\prime+\mathbf{p}+\frac{B^2}{2}\mathbf{I}-\mathbf{BB}
  \right)
  = 2 \rho \Omega v_y^\prime \hat{\mathbf{e}}_x
  + \left(q-2\right) \rho \Omega v_x^\prime \hat{\mathbf{e}}_y
  - \rho \Omega^2 z \hat{\mathbf{e}}_z, \\
  \frac{\partial \mathbf{B}}{\partial t}
  - \nabla \times \left( \mathbf{v}^\prime \times \mathbf{B}\right)
  = 0, \\
  \partial_t E_{ij}^\prime
  + \partial_k
  \left(\rho v_i^\prime v_j^\prime v_k^\prime
  + p_{ij} v_k^\prime + p_{ik} v_j^\prime + p_{jk} v_i^\prime
  + \mathcal{S}_{kij}^\prime + \mathcal{S}_{kji}^\prime\right)
  \nonumber \\ 
  =
  B_i v_k^\prime \partial_k B_j + B_j v_k^\prime \partial_k B_i
  - B_k v_i^\prime \partial_j B_k - B_k v_j^\prime \partial_i B_k
  \nonumber \\
  + v_i^\prime s_{g,j}^\prime
  + v_j^\prime s_{g,i}^\prime
  - \nu_g\left(p_{ij} - p_{g,ij}\right),
  \label{eqC:energy}
 \end{eqnarray}
 where quantities with primes are evaluated using $\mathbf{v}^\prime$
 rather than $\mathbf{v}$ itself, and
 \begin{eqnarray}
  \mathbf{s}_g^\prime =
   \left(
    \begin{array}{c}
     2\rho\Omega v_x^\prime \\
     -2\rho\Omega v_y^\prime \\
     -\rho\Omega^2 z
    \end{array}
   \right).
 \end{eqnarray}
 The gravitational and Coriolis forces are slightly modified from the
 original shearing box system.
 These equations are integrated in a standard manner.
 Note that Equations (\ref{eqC:linear_energy}) and (\ref{eqC:energy}) are
 reduced to Equations (51) and (55) in \cite{2010ApJS..189..142S},
 respectively, by taking their trace.

 \section{One-dimensional Riemann problem with the rotation term}
 \label{app:rotation}
 In order to clarify the role of the rotation term introduced in
 Section~\ref{subsec:rotation}, particularly on magnetic reconnection,
 let us reconsider the same one-dimensional Riemann problem as described
 in HHA16 by setting $\Omega_c=\left(B/B_0\right)\Omega_{c0}$ and
 $\nu_{g0}=0$.
 Here we briefly reproduce the initial setup.
 Let us consider a Harris-type current sheet at rest threaded by a
 uniform guide field and focus upon the one-dimensional cross section
 across the current sheet, which is taken as the $x$-direction.
 Then, the initial magnetic field is as follows:
 \begin{eqnarray}
  \mathbf{B}\left(x,t=0\right)
   = B_0 \cos\phi \tanh\left(x/L\right) \hat{\mathbf{e}}_y
   + B_0 \sin\phi \ \hat{\mathbf{e}}_z,
 \end{eqnarray}
 where $B_0$ is the magnetic-field strength in the lobe regions, $L$ is
 the half width of the initial sheet, and $\phi$ is fixed to $30^\circ$.
 The density and the isotropic thermal pressure are distributed so as to
 maintain thermal and dynamical equilibrium, respectively.
 Finally, reconnected flux $B_n = 5 \times  10^{-2} B_0$ is added in the
 $x$-direction, which triggers Petschek-type reconnection.
 
\begin{figure}[!ht]
  \centering
  \includegraphics[width=\textwidth]{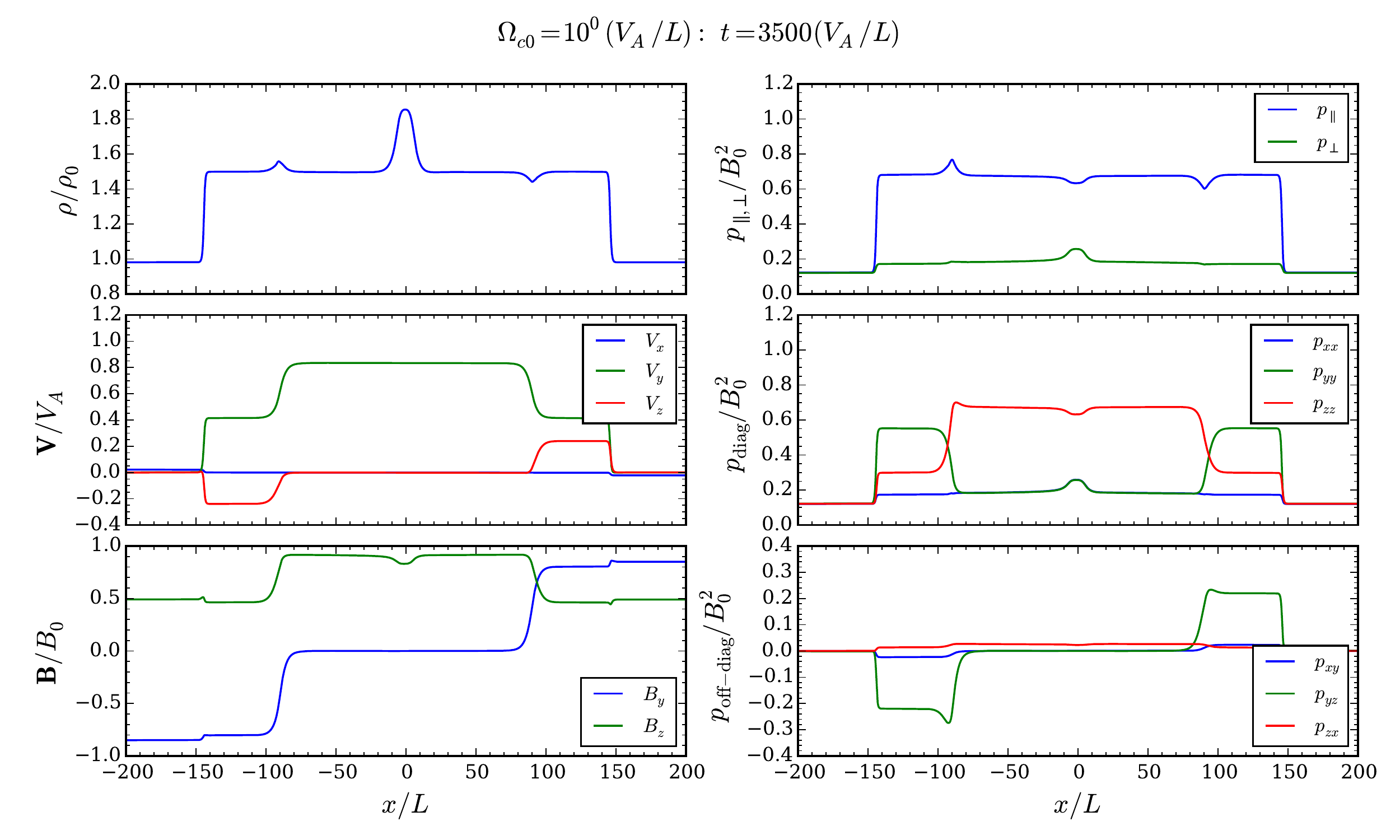}
  \caption{Results of the one-dimensional Riemann problem at
  $t=3,500$ including only the rotation term with $\Omega_{c0}=10^{0}$.}
  \label{fig:omg1.0e0}
 \end{figure}

\begin{figure}[ht]
  \centering
  \includegraphics[width=\textwidth]{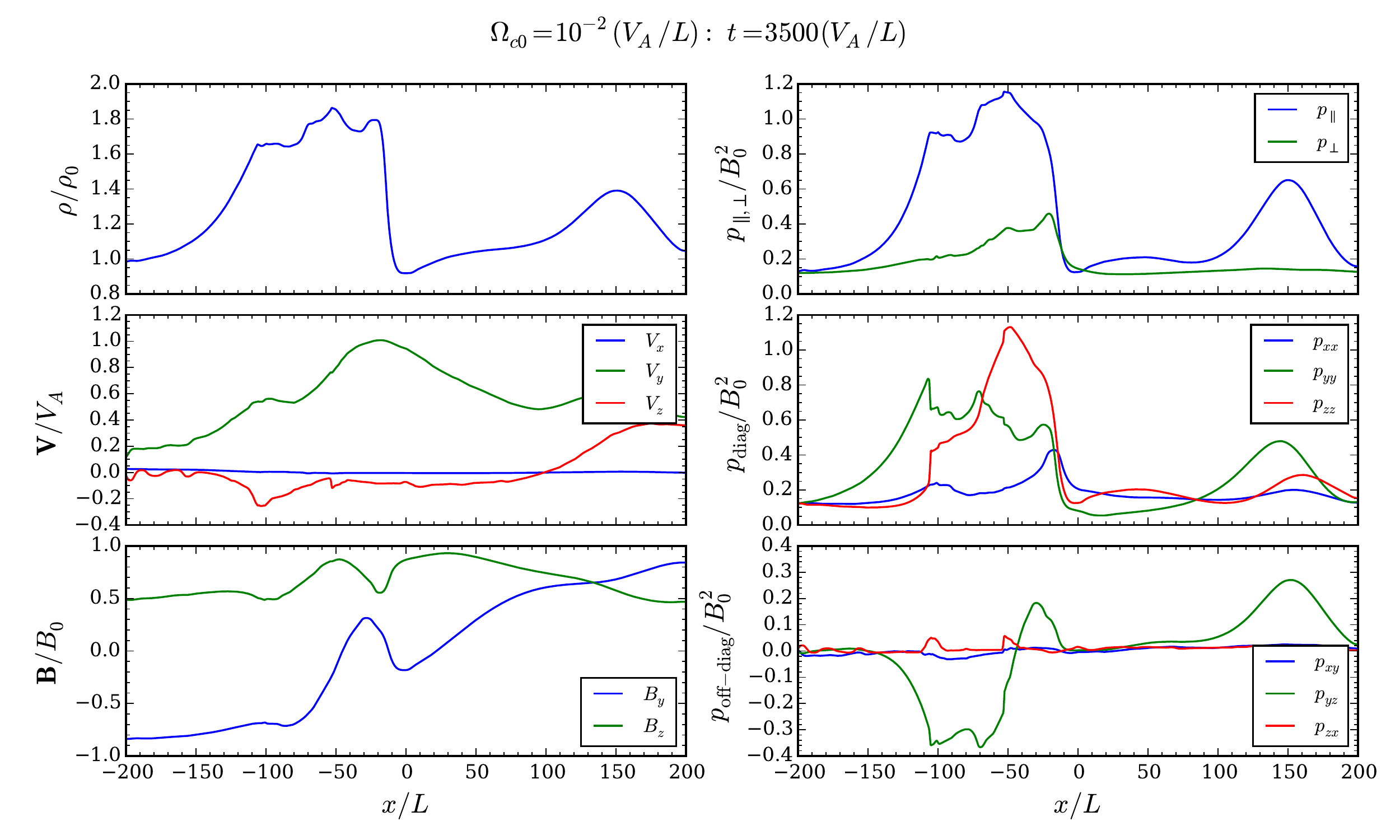}
  \caption{Results of the one-dimensional Riemann problem at
  $t=3,500$ including only the rotation term with $\Omega_{c0}=10^{-2}$.}
  \label{fig:omg1.0e-2}
 \end{figure}

\begin{figure}[ht]
  \centering
  \includegraphics[width=\textwidth]{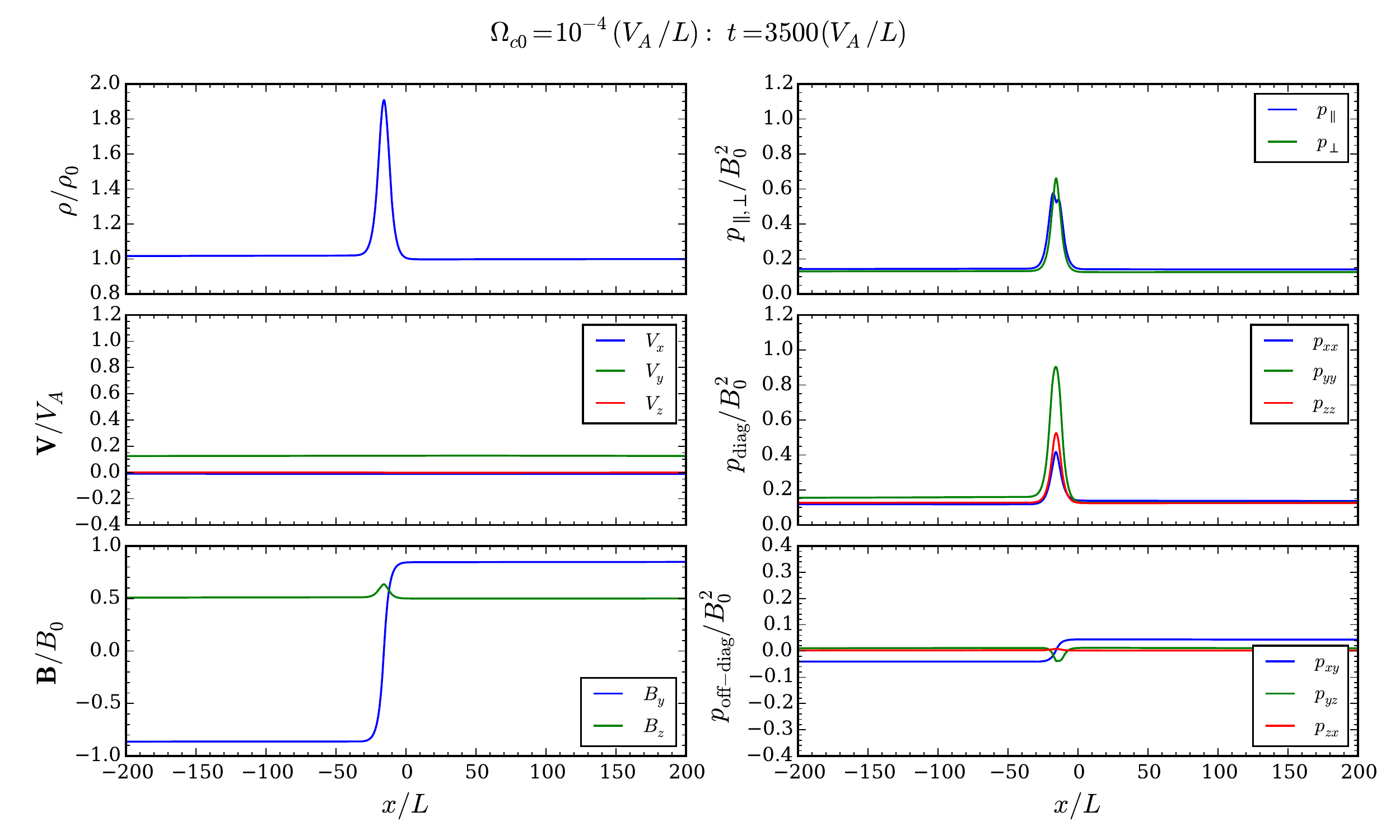}
  \caption{Results of the one-dimensional Riemann problem at
  $t=3,500$ including only the rotation term with $\Omega_{c0}=10^{-4}$.}
  \label{fig:omg1.0e-4}
 \end{figure}

 The snapshots adopting different values of the rotation frequency,
 i.e., $\Omega_{c0}L/V_A = 10^{0}$, $10^{-2}$, and $10^{-4}$ are shown
 in Figures~\ref{fig:omg1.0e0} to \ref{fig:omg1.0e-4}, respectively.
 By comparison between Figures~5 in HHA16 and Figure~\ref{fig:omg1.0e0},
 it is remarkable that when $\Omega_{c0}^{-1}$ takes the same order as
 the {\Alfven} transit time across the initial current sheet, the
 resultant structure of the reconnection layer is almost the same as
 that observed under the gyrotropic limit in spite of the absence of any
 explicit gyrotropization term.
 The resemblance to the gyrotropic case can be traced to
 redistribution of non-gyrotropic components generated at a certain
 gyrophase into a wide range of gyrophases.
 Fast rotation, therefore, tends to randomize non-gyrotropy, which
 effectively works like gyrotropization, although it is not guaranteed
 that non-gyrotropy vanishes eventually.
 The only major difference from the gyrotropic limit is the behavior of
 the rotational discontinuities around $\left|x/L\right| \sim 90$, where
 the responses of various quantities are anti-symmetric between leftward-
 and rightward-propagating waves.
 This symmetry breaking apparently arises from the assumption that the
 pressure is supported only by positive charges, i.e., $\Omega_{c}>0$,
 and thus reflects the direction of the local magnetic field.
  
 When the rotation is slower than the dynamical timescale, as shown in
 Figure~\ref{fig:omg1.0e-2} with $\Omega_{c0}=10^{-2}V_A/L$, the
 reconnection layer becomes quite complicated and highly asymmetric.
 It is no longer straightforward to understand the structure based on
 the propagation of isolated waves.
 Once the rotation frequency is further decreased by two orders of
 magnitude, however, the system shows no explosive reconnection, as
 shown in Figure~\ref{fig:omg1.0e-4}.
 This naturally occurs under the same mechanism described in
 HHA16 and Figure~6 there, and the assumption of positive $\Omega_{c}$
 causes a slight leftward migration of the contact discontinuity.
  

 \bibliography{reference}

\end{document}